\documentclass[twocolumn, 10pt, pra, aps, showpacs, reprint, amsmath, amssymb, superscriptaddress, nofootinbib, floatfix]{revtex4-1}

\usepackage{times}
\usepackage[pdftex]{graphicx} 

\begin{document}

\title{Sensitive control of broad-area semiconductor lasers by cavity shape}

\author{Kyungduk Kim}
\affiliation{Department of Applied Physics, Yale University, New Haven, CT 06520, USA}
\author{Stefan Bittner}
\affiliation{Chair in Photonics, LMOPS EA-4423 Laboratory, CentraleSup\'elec and Universit\'e de Lorraine, Metz 57070, France}
\author{Yuhao Jin}
\affiliation{Center for OptoElectronics and Biophotonics, School of Electrical and Electronic Engineering, School of Physical and Mathematical Science, and Photonics Institute, Nanyang Technological University, 639798 Singapore}
\author{Yongquan Zeng}
\affiliation{Center for OptoElectronics and Biophotonics, School of Electrical and Electronic Engineering, School of Physical and Mathematical Science, and Photonics Institute, Nanyang Technological University, 639798 Singapore}
\author{Stefano Guazzotti}
\author{Ortwin Hess}
\affiliation{School of Physics and CRANN Institute, Trinity College Dublin, Dublin 2, Ireland}
\author{Qi Jie Wang}
\affiliation{Center for OptoElectronics and Biophotonics, School of Electrical and Electronic Engineering, School of Physical and Mathematical Science, and Photonics Institute, Nanyang Technological University, 639798 Singapore}
\author{Hui Cao}
\email{hui.cao@yale.edu}
\affiliation{Department of Applied Physics, Yale University, New Haven, CT 06520, USA}

\begin{abstract}
The ray dynamics of optical cavities exhibits bifurcation points: special geometries at which ray trajectories switch abruptly between stable and unstable. A prominent example is the Fabry-Perot cavity with two planar mirrors, which is widely employed for broad-area semiconductor lasers. Such cavities support lasing in a relatively small number of transverse modes, and the laser is highly susceptible to filamentation and irregular pulsations. Here we demonstrate experimentally that a slight deviation from this bifurcation point (planar cavity) dramatically changes the laser performance. In a near-planar cavity with two concave mirrors, the number of transverse lasing modes increases drastically. While the spatial coherence of the laser emission is reduced, the divergence angle of the output beam remains relatively narrow. Moreover, the spatio-temporal lasing dynamics becomes significantly more stable compared to that in a Fabry-Perot cavity. Our near-planar broad-area semiconductor laser has higher brightness, better directionality and hence allows shorter integration times than an incandescent lamp while featuring sufficiently low speckle contrast at the same time, making it a vastly superior light source for speckle-free imaging. Furthermore, our method of controlling spatio-temporal dynamics with extreme sensitivity near a bifurcation point may be applied to other types of high-power lasers and nonlinear dynamic systems.
\end{abstract}

\maketitle

\renewcommand{\thesection}{\arabic{section}}

\section{Introduction}

The cavity is a crucial component determining the performance of a laser. There has been much effort in tailoring the cavity geometry to control lasing dynamics~\cite{siegman1990new}. Traditional cavity design is based on ray optics that tracks the propagation of optical rays inside a cavity. Thanks to the principle of ray-wave correspondence, the ray dynamics allows to predict many properties of the actual cavity resonances, such as their spatial structures and output patterns. Most solid state and gas laser cavities have two concave mirrors arranged in such a way that the axial orbit is stable~\cite{kogelnik1966laser,siegman1986lasers,hodgson2005laser}. In contrast, Fabry-Perot cavities with two planar facets have been widely adopted for semiconductor edge-emitting lasers. However, planar broad-area high-power semiconductor lasers are highly susceptible to filamentation and irregular pulsation due to strong nonlinear interactions of the optical field and the gain medium~\cite{fischer1996complex,hess1994spatio,hess1996maxwell,marciante1997spatio,marciante1998spatio,scholz2008measurement,arahata2015inphase}. 

\begin{figure}[t]
\centering
\includegraphics[width=\linewidth]{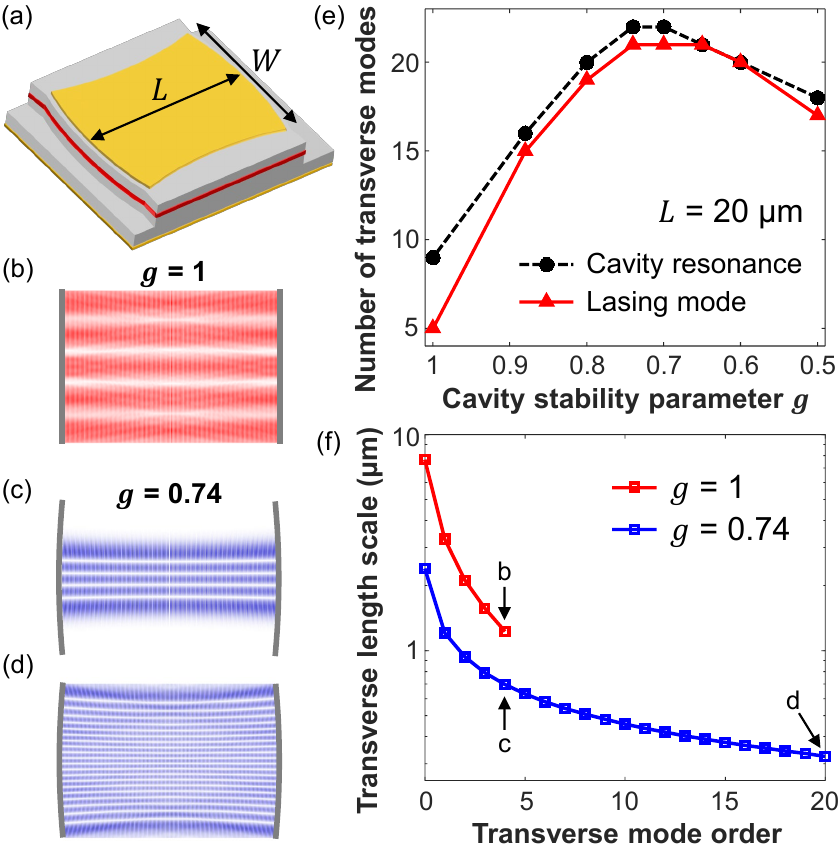}
\caption{Optical modes in a near-planar laser cavity. (a) Schematic of a near-planar broad-area semiconductor laser with cavity length $L$ and width $W$. (b) A passive mode of a planar cavity ($g = 1$) without mirror curvature, exhibiting a spatial profile extended over the entire facet. (c) A passive mode of the same transverse order in a near-planar cavity ($g = 0.74$), showing enhanced lateral confinement. (d) The profile of the highest-order transverse mode confined in the near-planar cavity ($g = 0.74$), featuring greatly decreased transverse wavelength. (e) The number of transverse modes in near-planar cavities. black dashed: passive resonances with a quality factor higher than $0.8 \, Q_{\rm max}$. red solid: lasing modes at two times the lasing threshold in the presence of gain competition. (f) Reduction of the transverse characteristic length scale of the passive modes (on a logarithmic scale). Each curve stops at the highest-order transverse lasing mode calculated in (e). The modes in (b)-(d) are indicated with arrows.}
\label{fig1}
\end{figure}

An interesting aspect of the planar Fabry-Perot cavity [see Fig.~\ref{fig1}(b)] with profound impact on the lasing dynamics is that it is situated at the bifurcation between stable cavities with concave mirrors [Fig.~\ref{fig1}(c)] and unstable cavities with convex mirrors. One way of suppressing the semiconductor laser instabilities is to destabilize the cavity ray dynamics by tilting the planar facet~\cite{salzman1985tilted,lang1986modal} or changing it to a convex shape~\cite{salzman1985unstable,salzman1986confocal,tilton1991high,biellak1997reactive}. Such cavities lase only in the fundamental mode, which stabilizes the temporal dynamics~\cite{adachihara1993spatiotemporal}. However, at high pump powers, additional transverse modes can lase nonetheless, and their nonlinear interactions with the gain medium brings back filaments and pulsations. 

Instead of considering unstable cavities, we have moved deeply into the regime of stable cavities for highly multimode lasing in previous works \cite{kim2021massively}. By replacing the flat mirrors with concave ones of large curvature, high-order transverse modes are well confined in near-concentric cavities~\cite{biellak1995lateral,kim2019electrically}. Their small transverse wavelength prevents filament formation and mitigates spatio-temporal instabilities~\cite{kim2019electrically, kim2021massively}. In contrast, such stable cavities confine ray trajectories with a large range of propagation directions, leading to strongly divergent far-field emission~\cite{kim2019electrically}. For efficient collection of the laser emission, it is desirable to also attain high output directionality by staying closer to the Fabry-Perot geometry while maintaining low spatial coherence and stable lasing dynamics. 

Here we investigate broad-area edge-emitting semiconductor laser performance as a function of the resonator geometry in the vicinity of the bifurcation at the planar cavity geometry. We experimentally demonstrate that a tiny modification of the Fabry-Perot cavity has a profound impact on the lasing dynamics and spatial coherence. As we slightly curve the two end facets to form a near-planar stable cavity with concave mirrors, the spatial structures of cavity resonances are strongly modified, which in turn alters their nonlinear interactions with the gain medium (GaAs quantum well). Consequently, the spatio-temporal stability of the lasing dynamics in near-planar cavities is greatly improved. Such a simple scheme of mitigating instabilities will facilitate the stabilization of high-power broad-area semiconductor lasers for applications in material processing~\cite{li2000advances} and laser pumping~\cite{shigihara1991high}, as well as biomedical applications~\cite{muller2013diode}. 

Moreover, the number of transverse lasing modes drastically increases compared to that in the planar Fabry-Perot cavity, resulting in a sharp drop of the spatial coherence of the emission. At the same time, the output beam has a far-field divergence angle notably smaller than that of stable cavities with strongly-curved facets~\cite{kim2019electrically}. 
This combination of sufficiently low spatial coherence and relatively good emission directionality makes our laser an ideal illumination source for speckle-free full-field imaging~\cite{redding2012speckle,cao2019complex}.

\section{Near-planar cavity}

\subsection{Ray dynamics}

Starting with the Fabry-Perot cavity, we gradually change the planar mirrors to concave ones, while keeping the distance $L$ between the mirrors and the cavity width $W$ constant. The cavity stability parameter is given by $g = 1 - L/R$, where $R$ is the radius of curvature of the end mirrors. In a Fabry-Perot cavity with $g = 1$, an optical ray is trapped only if it propagates parallel to the cavity axis (perpendicular to the end mirrors). When the propagation direction of a ray deviates from the cavity axis, it runs laterally out of the cavity after a few round trips [see Supplementary Material]. 

With concave mirrors at both ends, $g$ becomes less than 1, and the ray dynamics changes substantially. In a near-planar cavity with even slightly curved mirrors [Fig.~\ref{fig1}(a)], the axial orbit becomes stable, and additional trajectories propagating with a slight angle with respect to the cavity axis remain confined laterally.

\subsection{Cavity resonances} 

The dramatic change of the ray dynamics from planar to near-planar cavities has a strong influence on the spatial structure of the cavity resonances (i.e., solutions of the wave equation of the passive cavity). Due to the lack of lateral confinement in a Fabry-Perot cavity, the cavity resonances extend laterally across the entire end facets [Fig.~\ref{fig1}(b)]. As the higher-order transverse modes exhibit larger transverse wavevector components $k_\perp$, their lateral leakage is stronger, and their quality ($Q$) factor is lower. In contrast, the lateral confinement of rays by concave mirrors reduces the transverse width of the cavity resonance, as seen in Fig.~\ref{fig1}(c). The existence of confined trajectories in the vicinity of the cavity axis greatly enhances the $Q$ factor of high-order transverse modes. Therefore, even a near-planar cavity can feature a relatively large number of transverse modes [Fig.~\ref{fig1}(e)]. 

The existence of high-order transverse modes and their lateral confinement in a stable cavity lead to a sharp drop in the characteristic length scale $\xi$ of the optical intensity variation in the transverse direction. This has a profound impact on the nonlinear interactions between the cavity modes and the gain medium~\cite{bittner2018suppressing,kim2021massively}. $\xi$ is given by the full-width at half-maximum of the transverse intensity correlation function. Since $\xi$ varies in the longitudinal direction for $g \neq 1$, we average its value along the cavity axis. Higher-order transverse modes have smaller $\xi$. In a Fabry-Perot cavity with a GaAs quantum well (gain medium), only low-order transverse modes are confined, and intensity variations on the scale $\xi \gg \lambda$ result in local carrier-induced refractive index changes due to spatial hole burning. Optical lensing and self-focusing effects lead to the formation of spatial filaments, which are inherently unstable and cause irregular pulsations. 

In a stable cavity, the $\xi$ of high-order transverse modes can be sufficiently small so local refractive index variations cannot focus light and create a filament. Also the spatial modulation of the refractive index on such short scales supersedes and disrupts the large lenses induced by lower-order transverse modes, thus preventing filamentation. Therefore, an efficient way of reducing the spatio-temporal instability in a broad-area semiconductor quantum well laser is to minimize $\xi$ via a stable cavity \cite{kim2021massively}.

\subsection{High-Q modes} 

To maintain a relatively narrow angular spread of the far-field emission, we optimize the stable cavity geometry in the vicinity of the planar cavity. In particular, we maximize the number of transverse lasing modes within the range $0.5 \leq g \leq 1$. Since the lasing modes need to have high $Q$ factors, we calculate the high-$Q$ transverse modes of passive cavities. Because the GaAs quantum well has preferential gain for transverse electric (TE) polarization, we calculate the modes with the electric field in the plane of incidence (p-polarization). The simulated cavities have the same aspect ratio $L/ W = \sqrt{2}$ as the experimental ones, but are $L$ = 20 $\mu$m long. The refractive index $n$ = 3.37 in the cavity is equal to the effective index of the fundamental TE mode guided in the vertical direction of an GaAs/AlGaAs epiwafer. The wavelength range of numerical simulations is centered around 800~nm, which matches the gain spectrum of the GaAs quantum well. 

The fundamental transverse mode has the highest quality factor $Q_{\rm max}$, which determines the lasing threshold. $Q_{\rm max}$ depends on the cavity length $L$ and mirror reflectivity. It barely changes with $g$, thus curving the end facets has little effect on the lasing threshold. However, the $Q$ factors of the high-order transverse modes are greatly improved in a stable cavity with concave mirrors, leading to a drastic increase in the number of lasing modes.

We show the number $M_h$ of transverse modes with $Q > 0.8 \, Q_{\rm max}$ in Fig.~\ref{fig1}(e). As $g$ decreases from 1, $M_h$ increases rapidly and reaches its maximum at $g$ = 0.7, then decreases. While the sharp rise results from the better lateral confinement of high-order transverse modes, the subsequent drop of $M_h$ is caused by the reduced reflectivity for rays with non-perpendicular incidence (see Supplementary Material). With increasing mirror curvature, the angles of incidence $\theta_i$ at the semiconductor-air interface increase, and for TE polarization the reflectivity decreases for $\theta_i$ going from 0 towards the Brewster angle~\cite{kim2019electrically}. Therefore, the number of high-$Q$ transverse modes is maximal for a near-planar cavity. 

\subsection{Number of lasing modes}

Due to their competition for gain, not all high-$Q$ modes will lase eventually. Taking into account the spatial overlap of these modes, we compute the number $M_l$ of transverse lasing modes using the Single-Pole-Approximation Steady-state \textit{ab-initio} Lasing Theory (SPA-SALT)~\cite{ge2010steady}. Figure~\ref{fig1}(e) compares $M_l$ at two times the lasing threshold to $M_h$. The difference between $M_h$ and $M_l$ reflects the strength of modal competition for gain. In the Fabry-Perot cavity of $g = 1$, only 5 transverse modes manage to lase among 9 high-$Q$ passive modes. In contrast, 21 out of 22 high-$Q$ transverse modes lase in a near-planar cavity of $g$ = 0.7. Thus, slightly curving the end mirrors raises the number of transverse lasing modes by a factor of 4. 

The transverse length scale $\xi$ of the intensity variation is plotted for every transverse lasing mode in Fig.~\ref{fig1}(f). For the same transverse mode order, $\xi$ is much smaller for $g$ = 0.74 than for $g$ = 1 due to better lateral confinement. In addition, more higher-order transverse modes lase, further reducing $\xi$. The shortest $\xi$, corresponding to the highest-order transverse lasing mode, drops 4 times from $g = 1$ to $0.74$. The dramatic decrease of the transverse intensity variation length scale prevents the formation of carrier-induced optical lenses. It suppresses self-focusing that would lead to filamentation and instabilities. As a result, the near-planar cavity feature the improved laser stability, similar to a cavity with chaotic ray dynamics~\cite{bittner2018suppressing}.

\section{Highly multimode lasing}

We fabricate near-planar laser cavities on a GaAs quantum well epiwafer. The cavity structure is defined by dry etching with an etch depth of 3.5 $\mu$m (see Supplementary Material). While the radius of curvature $R$ of the two end facets is varied from device to device, the longitudinal cavity length $L$ = 400 $\mu$m and transverse cavity width $W$ = 283 $\mu$m are kept constant. The aspect ratio $L/W = \sqrt{2}$ is consistent with the simulations in Fig.~\ref{fig1}, even though the cavity size is 20 times larger. To reduce sample heating, we operate the lasers at room temperature with a pulse duration of 2 $\mu$s and a repetition rate lower than 1 Hz. All devices with different $g$ have similar lasing thresholds of approximately 0.5 A, corresponding to a current density of 0.5 kA/cm$^2$ (see Supplementary Material).

\subsection{Number of transverse lasing modes}

\begin{figure}[t]
\centering
\includegraphics[width=\linewidth]{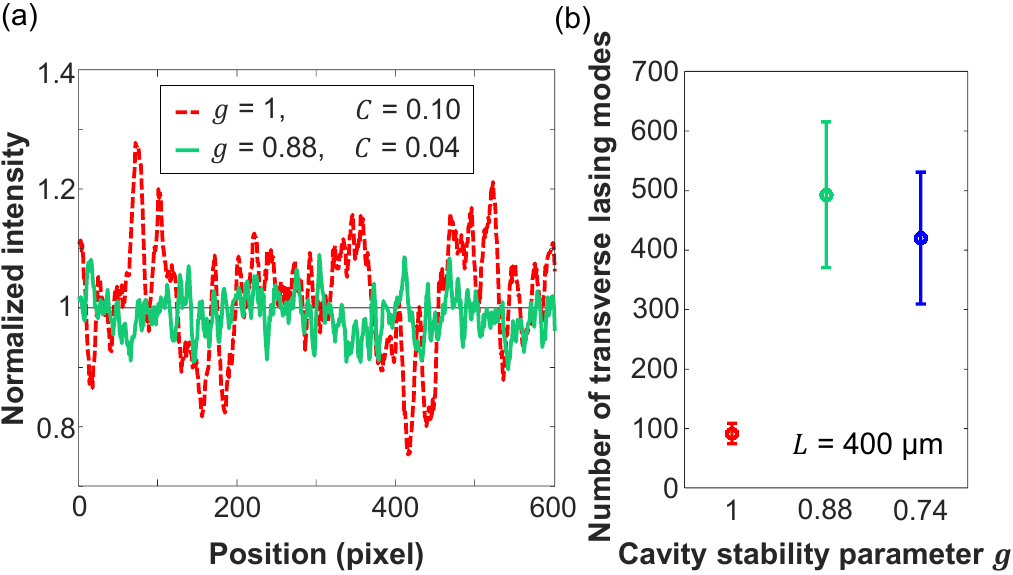}
\caption{Number of transverse lasing modes in near-planar cavities. (a) Measured intensity fluctuations of far-field speckles created by a line diffuser illuminated with laser emission from planar (red dashed) and near-planar (green solid) cavities. The speckle contrast $C$ is reduced from 0.1 at $g = 1$ to 0.04 at $g = 0.88$. (b) Number of transverse lasing modes $M_l$, estimated from $C$, at two times of the lasing threshold.  All laser cavities have length $L$ = 400 $\mu$m (20 times longer than the simulated cavities in Fig.~\ref{fig1}). $M_l$ increases sharply as $g$ decreases from 1. The error bars denote variations among multiple fabricated devices of the same geometry $g$.}
\label{fig2}
\end{figure}

We measure the number of transverse lasing modes in cavities of different $g$. The laser emission passes through a diffuser and creates a speckle pattern in the far field. The number of transverse lasing modes $M_l$ is estimated as $1/C^2$, where $C$ is the speckle intensity contrast~\cite{goodman2007speckle, redding2015low, kim2019electrically}. Because the edge emission from the laser contains multiple transverse modes in the horizontal direction and a single guided mode in the vertical direction (perpendicular to the wafer), a line diffuser (RPC Photonics, EDL-20) is used and the far-field speckle intensity variation in the horizontal direction is recorded, as shown in Fig.~\ref{fig2}(a).  

The speckle intensity contrast $C$ decreases with decreasing $g$, indicating an increase in the number of transverse lasing modes $M_l$. For $g$ = 1, the number of transverse lasing modes, averaged over multiple devices, is approximately 100. 
Once $g$ is reduced slightly to 0.88, $M_l$ is enhanced 5 times to about 500. Such rapid increase of $M_l$ is consistent with the numerical simulation in Fig.~\ref{fig1}(e). Further reducing $g = 0.74$ does not increase $M_l$ any more, in contrast to the numerical results in Fig.~\ref{fig1}(e). We attribute this difference to the spatially inhomogeneous current injection in our devices, which modifies the number of transverse lasing modes $M_l$ (see Supplementary Material). With spatially homogeneous pumping, we expect $M_l$ to be higher, particularly for $g = 0.74$. We also note that the number of transverse lasing modes in the near-planar cavities is close to the previously reported value for a near-concentric ($g = -0.74$) cavity~\cite{kim2019electrically}. Therefore, even minimal curvature of the end facets is sufficient to obtain a large number of transverse lasing modes and reduce the spatial coherence.

\subsection{Divergence of far-field emission}

\begin{figure}[t]
\centering
\includegraphics[width=\linewidth]{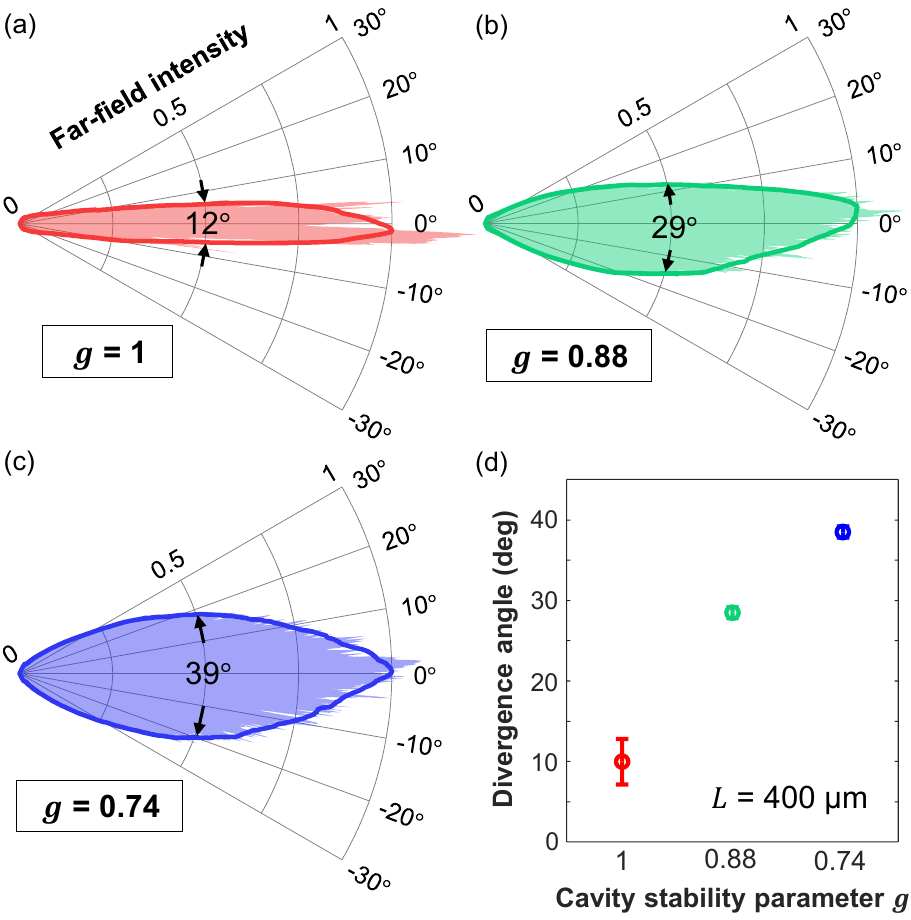}
\caption{Emission directionality of near-planar cavity lasers. (a-c) Measured far-field emission patterns $I(\theta)$ from cavities of (a) $g$ = 1, (b) 0.88, and (c) 0.74. The shaded area represents the measured intensity profile, and the solid line denotes the smoothed profile with the maximal value normalized to 1. The angular full-width at half maximum $\Delta \theta$ is indicated by arrows. (d) Divergence angles $\Delta \theta$ as function of $g$. All cavities have the same dimensions as those in Fig.~\ref{fig2}. The error bars denote variations among different cavities with the same $g$.}
\label{fig3}
\end{figure}

As more transverse modes lase, the divergence angle of the total emission increases. We measure the far-field emission pattern $I(\theta)$ in the horizontal direction, as shown in Fig.~\ref{fig3}. The filaments in a planar cavity of $g = 1$ makes $I(\theta)$ asymmetric and irregular [Fig.~\ref{fig3}(a)]. The divergence angle $\Delta \theta$ is estimated by the full-width at half-maximum of the smoothed distribution of $I(\theta)$, and it equals 12$^\circ$. With curved end facets, $\Delta \theta$ increases to 29$^\circ$ for $g = 0.88$ [Fig.~\ref{fig3}(b)], and further to 39$^\circ$ for $g = 0.74$ [Fig.~\ref{fig3}(c)]. We note that the measured far-field emission patterns are narrower than the simulated ones due to spatial inhomogeneity of current injection (see Supplementary Material). While the increase of lateral divergence angle in the near-planar cavities [Fig.~\ref{fig3}(d)] is expected, the emission directionality is significantly improved compared to the near-concentric cavity~\cite{kim2019electrically} with $g=-0.74$ and $\Delta \theta = 70^{\circ}$. Moreover, the lateral divergence of emission from the near-planar cavity is comparable to the vertical divergence of the edge emitting laser. Therefore, the output beam is approximately circular and thus compatible with standard collection optics.

\section{Spatio-temporal dynamics}

\begin{figure*}[t]
\centering
\includegraphics[width=\textwidth]{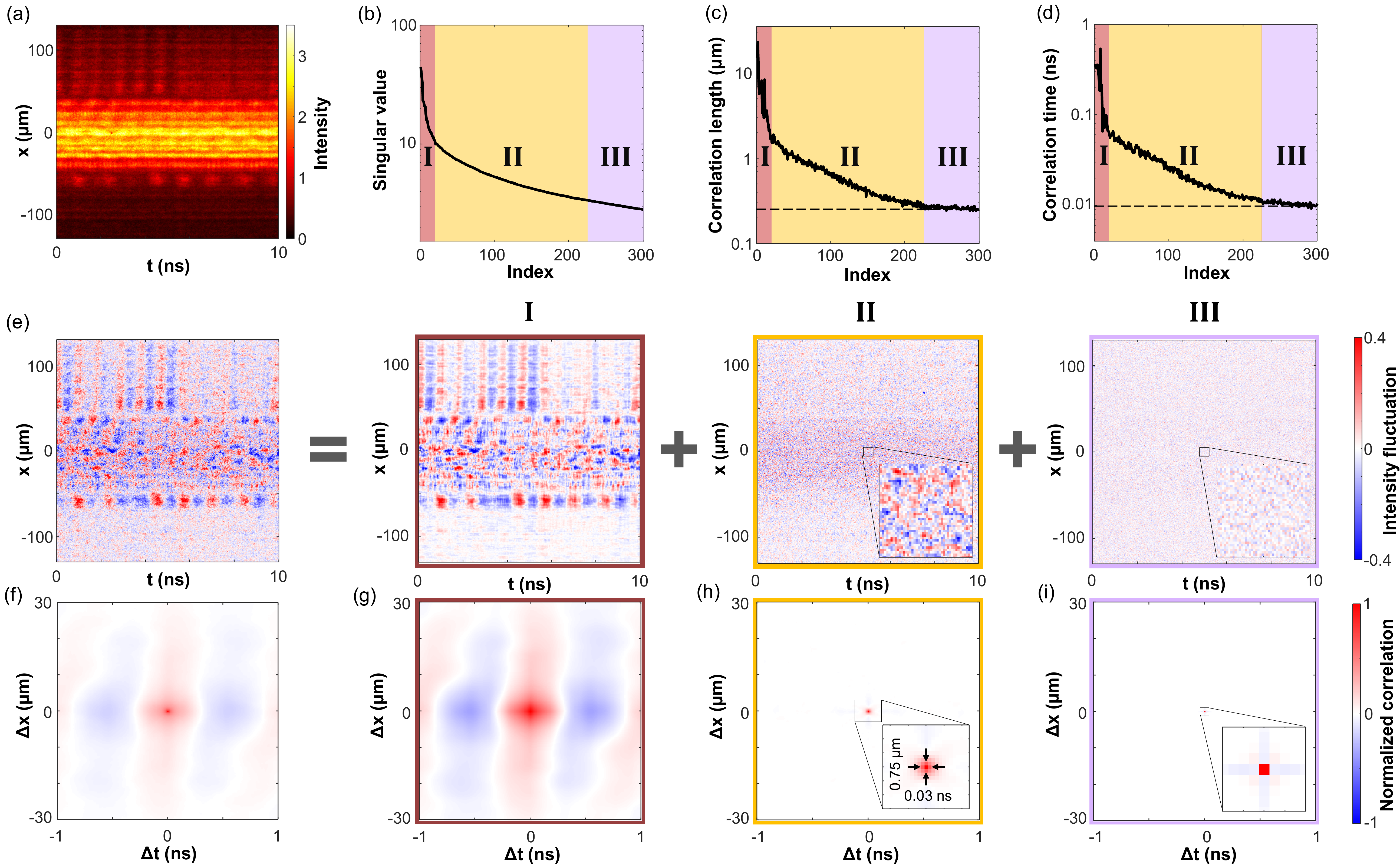}
\caption{Separating spatio-temporal intensity fluctuations of different origin by singular value decomposition (SVD). (a) Streak image of emission intensity $I(x,t)$ on one facet of a Fabry-Perot cavity ($g = 1$). The pump current is two times the lasing threshold. (b)~Singular values obtained by SVD of intensity fluctuations $\delta I(x,t) = I(x,t) - \langle I(x,t) \rangle_t$. (c) Correlation lengths and (d) correlation times of singular vectors. The horizontal dashed lines denote the size of a single pixel in space and time of the streak image. The singular vectors are  categorized into three groups I, II, and III, which are marked by red, yellow and purple, respectively. (e) $\delta I(x,t)$ is the sum of intensity fluctuations caused by (I) filaments, (II) mode beating, and (III) detection noise. (f) The spatio-temporal correlation function of $\delta I(x,t)$ features both short- and long-range correlations. (g-i) Spatio-temporal correlation functions of intensity fluctuation for (g) group I (h) II, and (i) III, exhibiting distinct correlation scales in space and time. Insets: close-up around the origin. The numbers in (h) are the spatial and temporal correlation widths.}
\label{fig4}
\end{figure*}

Next we investigate the lasing dynamics of near-planar cavities and compare to the planar cavity. The emission intensity on one facet of the cavity is imaged by a $\times$20 objective lens onto the entrance slit of a streak camera (Hamamatsu C5680 with a fast sweep unit M5676). A single streak image covers a time window of 10 ns with a temporal resolution of about 30 ps. Figure~\ref{fig4}(a) shows exemplary spatio-temporal intensity fluctuations of the laser emission from a planar cavity of $g = 1$. Such fluctuations comprise three different processes: (i) spatial filaments and their pulsations; (ii) spatio-temporal interference of  transverse and longitudinal lasing modes; (iii) photo-detection noise of the streak camera. To separate the three fluctuation processes, we resort to the fact that they feature distinct spatial and temporal scales.

\subsection{Separation of different fluctuations}

After normalizing the measured intensity $I(x,t)$ by the average $\langle I(x,t) \rangle_{x,t}$, we conduct the singular value decomposition (SVD) of the intensity fluctuation $\delta I(x,t) = I(x,t) - \langle I(x,t) \rangle _t$:
\begin{equation}
\delta I(x,t) = \sum_{\alpha} s_{\alpha} u_{\alpha}(x) v_{\alpha}(t)
\label{eq1}
\end{equation}
where $s_{\alpha}$ are singular values, $u_{\alpha}(x)$ and $v_{\alpha}(t)$ are the spatial and temporal singular vectors, respectively, and $\alpha$ denotes the index. The singular values $s_{\alpha}$ are arranged from high to low, and they represent the contributions of the $\alpha$-th singular vector to the intensity fluctuation $\delta I(x,t)$. As shown in Fig.~\ref{fig4}(b), $s_{\alpha}$ first drops sharply with increasing $\alpha$, then decays more slowly for $\alpha > 20$. Hence, the first few singular vectors with large singular values dominate $\delta I(x,t)$. 

To distinguish the singular vectors, we analyze their characteristic spatial and temporal scales. The correlation functions of $u_{\alpha}(x)$ and $v_{\alpha}(t)$ are defined as
\begin{equation}
\begin{aligned}
C_{\alpha}(\Delta x) &= \langle u_{\alpha}(x)u_{\alpha}(x + \Delta x) \rangle_{x} \\
C_{\alpha}(\Delta t) &= \langle v_{\alpha}(t)v_{\alpha}(t + \Delta t) \rangle_{t}.
\end{aligned}
\label{eq2}
\end{equation}
The correlation length $l_{\alpha}$ is extracted from the full-width at half-maximum of $C_{\alpha}(\Delta x)$, and the correlation time $\tau_{\alpha}$ from $C_{\alpha}(\Delta t)$. As shown in Figs.~\ref{fig4}(c) and (d), both $l_{\alpha}$ and $\tau_{\alpha}$ decrease rapidly with increasing $\alpha$, till $\alpha$ reaches 20. Then they switch to a more gradual decay and eventually level off. When $\alpha$ exceeds 220, both spatial and temporal correlation scales are equal to the single pixel size of the streak image. 

Since $s_{\alpha}$, $l_{\alpha}$ and $\tau_{\alpha}$ exhibit similar dependency on $\alpha$, we separate the singular vectors into three groups, denoted as I, II, and III in Figs.~\ref{fig4}(b)-(d). Then the spatio-temporal intensity fluctuation of each group is reconstructed by 
\begin{equation}
\begin{aligned}
\delta I_\textrm{R}(x,t) &= \sum_{\alpha \in \textrm{R}} s_{\alpha} u_{\alpha}(x) v_{\alpha}(t)
\end{aligned}
\label{eq3}
\end{equation}
where R is one of the three groups $\mathrm{I}$, $\mathrm{II}$, and $\mathrm{III}$. The total intensity fluctuation is $\delta I(x,t) = \delta I_\textrm{I}(x,t) + \delta I_\textrm{II}(x,t) + \delta I_\textrm{III}(x,t)$. In Fig.~\ref{fig4}(e), $\delta I_\textrm{I}(x,t)$, $\delta I_\textrm{II}(x,t)$, and $\delta I_\textrm{III}(x,t)$  display different spatial and temporal scales. To quantify the difference, we compute the spatio-temporal correlation function for each group,
\begin{equation}
C_\textrm{R}(\Delta x,\Delta t) = \langle \delta I_\textrm{R}(x,t)\delta I_\textrm{R}(x+ \Delta x,t+ \Delta t)\rangle_{x,t}.
\label{eq4}
\end{equation}
The spatial and temporal widths of $C_\textrm{R}(\Delta x,\Delta t)$ give the correlation lengths and times for every group.

The first group $\delta I_\textrm{I}(x,t)$ features strong intensity fluctuations on length scales from several to tens of micrometers, and a time scale of the order of 0.1 nanoseconds. Such scales are consistent with the typical size of spatial filaments and their oscillation frequencies. $C_\textrm{I}(\Delta x,\Delta t)$ in Fig.~\ref{fig4}(g) reveals long-range spatio-temporal correlations as a result of filament motion and pulsation. 

The second group $\delta I_\textrm{II}(x,t)$ features fluctuations on much shorter spatial and temporal scales. As seen in Fig.~\ref{fig4}(e), $\delta I_\textrm{II}(x,t)$ is stronger in the middle of the cavity (around $x$ = 0), where the original emission intensity $I(x,t)$ in Fig.~\ref{fig4}(a) is stronger. It implies that the group II also originates from the laser emission, more precisely, from the spatio-temporal interference of all lasing modes. $C_\textrm{II}(\Delta x,\Delta t)$ in Fig.~\ref{fig4}(h) exhibits only local correlations of the intensity fluctuations. Both the correlation length of 0.75 $\mu$m and correlation time of 0.03 ns are limited by the resolution of our photodetection.

For the third group, $\delta I_\textrm{III}(x,t)$ is uniformly spread over the entire range of the streak image. It indicates the fluctuation is not due to the laser emission, but from the noise generated by the imaging apparatus. This is confirmed by $C_\textrm{III}(\Delta x,\Delta t)$ in Fig.~\ref{fig4}(i), which shows the spatio-temporal correlation scales equal to the pixel size of the image. Therefore, $\delta I_\textrm{III}(x,t)$ represents the detection noise that fluctuates on the scale of a single pixel. 

\begin{figure*}[t]
\centering
\includegraphics[width=\textwidth]{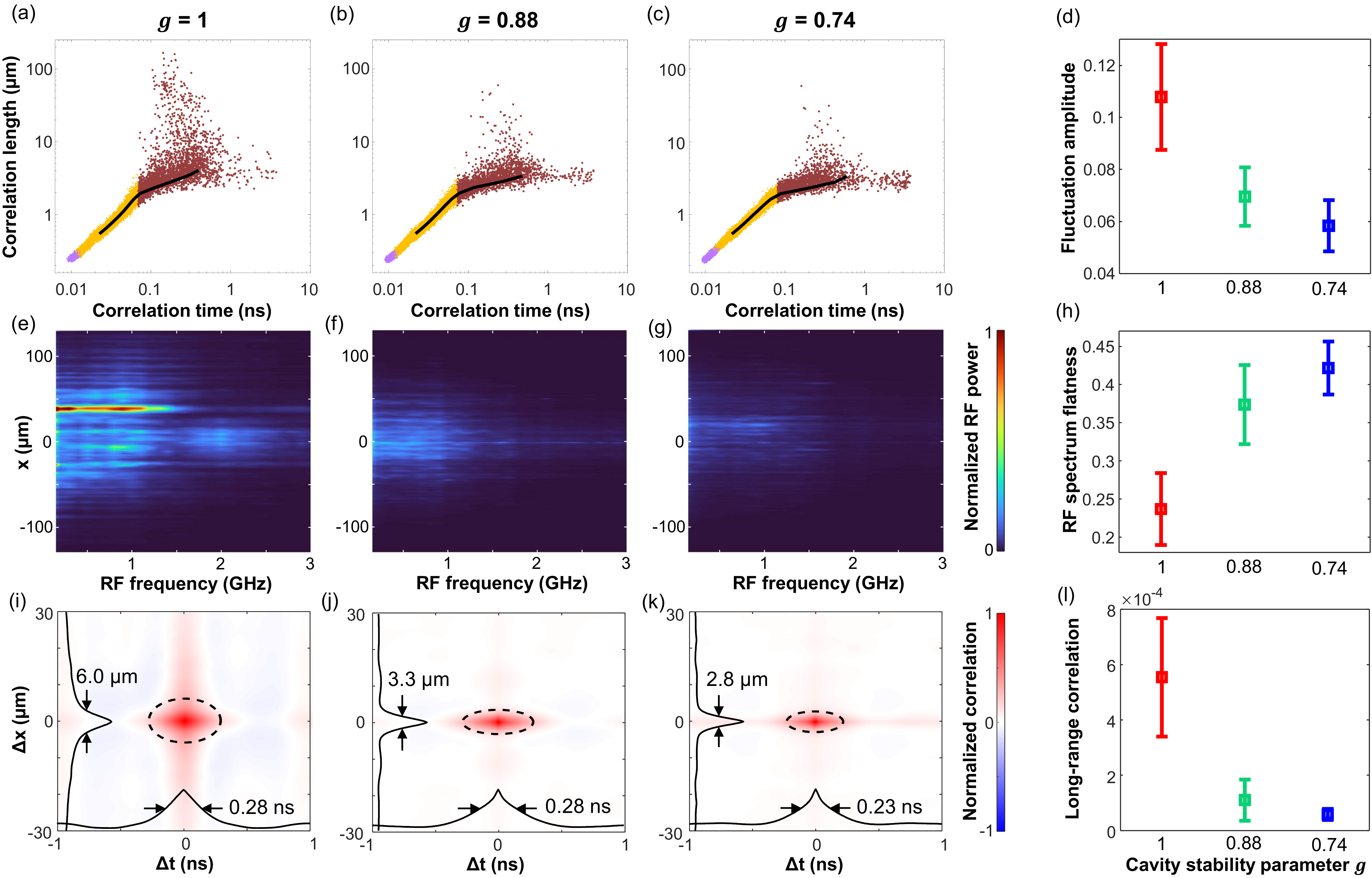}
\caption{Spatio-temporal dynamics of near-planar cavity lasers. (a-c) Scatter plots of correlation lengths and times for all singular vectors of intensity fluctuations in laser cavities with (a) $g$ = 1, (b) 0.88, and (c) 0.74. The black line denotes the average correlation time at a fixed correlation length, and its slope changes suddenly at the correlation time of 0.07 ns. The three groups of singular vectors with different origin are marked by (I) red, (II) yellow and (III) purple, respectively. (d) The amplitude of spatio-temporal intensity fluctuations caused by filaments (group I). The error bars denote variations among 5 fabricated cavities of the same $g$. (e-g) Spatially resolved RF spectra for (e) $g$ = 1, (f) 0.88, and (g) 0.74, showing a clear reduction of oscillatory power as $g$ decreases. (h) Flatness of the spatially-averaged RF spectrum for group I. The increase in flatness indicates the suppression of RF peaks corresponding to intensity oscillations. (i-k) Spatio-temporal correlation functions of the intensity fluctuations for group I, $C_I(\Delta x, \Delta t)$, in laser cavities of (i) $g$ = 1, (j) 0.88, and (k) 0.74. The black solid lines are $C_I(\Delta x, 0)$ and $C_I(0, \Delta t)$. Their widths give the correlation length and time. The ellipse (black dashed line) divides the regions of local and nonlocal correlations, and its semi-axes are equal to correlation length and time. (l) Long-range spatio-temporal correlations in (i-k), given by the averaged magnitude of the correlation outside the ellipse.}
\label{fig5}
\end{figure*}

\subsection{Spatio-temporal instabilities}

The SVD can efficiently separate intensity fluctuations of different scales and origins. In the example of Fig.~\ref{fig4}, the three groups contain singular vectors with consecutive indices $\alpha$. In general, the singular vectors in each group have similar correlation scales, but not necessarily adjacent indices. Below we present a systematic way to group the singular vectors based on their correlation scales.

We acquire an ensemble of spatio-temporal intensity distributions from the same laser cavity. Within a 2-$\mu$s-long current pulse, we take 161 consecutive streak images, each covering a 10 ns window. The transient regime at the beginning of the current pulse is excluded. We perform the SVD on every streak image and compute the correlation length $l_\alpha$ and time $\tau_\alpha$ for each singular vector. A scatter plot of $l_\alpha$ and $\tau_\alpha$ for all singular vectors in all streak images is presented in Fig.~\ref{fig5}(a). The black line denotes the average of correlation times for a given correlation length. Its slope changes suddenly at the correlation time of 0.07 ns, which is consistent with the boundary between group I and II in Fig.~\ref{fig4}(d). Moreover, the singular vectors with correlation times longer than 0.07 ns exhibit a large variation of the correlation length. Therefore, they belong to group I that originates from the filaments. For each time window, we sum all singular vectors with $\tau_\alpha$ > 0.07 ns of a single streak image to reconstruct $\delta I_\textrm{I}(x,t)$ (see Supplementary Material).

\subsection{Relative intensity fluctuation}

Using the same method, we investigate the spatio-temporal lasing dynamics in the near-planar cavities. Figures~\ref{fig5}(b) and (c) are the scatter plots of spatial and temporal correlation scales for $g$ = 0.88 and 0.74. Compared to $g = 1$ [Fig.~\ref{fig5}(a)], the singular vectors with correlation lengths exceeding 10 $\mu$m are notably fewer, indicating slight curving of the end facets leads to a reduction of filamentation.

For a quantitative comparison, we introduce a measure of the spatio-temporal instability. It can be shown that the square of a singular value is proportional to the fluctuation power of the corresponding singular vector (see Supplementary Material). Thus the root-mean square of the singular values $s_{\alpha}$ in group I for all time windows yields a measure for the strength of spatio-temporal instabilities, which is called the fluctuation amplitude $p$.

Figure~\ref{fig5}(d) compares $p$ for different cavity shapes. We average over 5 fabricated lasers for each $g$, and the error bars reflect the device-to-device variation. The mean fluctuation amplitude $p$ for $g$ = 0.74 is about half of that for $g = 1$. Therefore, the spatio-temporal dynamics becomes much more stable in the near-planar cavity lasers.

\subsection{RF spectrum}

Next, we perform the Fourier transform of $\delta I_\textrm{I}(x,t)$ to obtain the spatially-resolved radio-frequency (RF) spectrum for group I:
\begin{equation}
P(x,f) = \langle |\mathcal{F}\{\delta I_\textrm{I}(x,t)\}|^2 \rangle
\label{eq8}
\end{equation}
where $\mathcal{F}$ denotes the temporal Fourier transform, and $\langle \cdot \rangle$ represents the average over different time windows. With the normalization $\langle I(x,t) \rangle_{x,t} = 1$, it is possible to compare the RF power from different cavities. In Fig.~\ref{fig5}(e), $P(x,f)$ for $g$ = 1 displays strong oscillations at a few GHz. Such oscillations become much weaker for $g$ = 0.88 and 0.74 in Figs.~\ref{fig5}(f) and (g). 

Suppression of intensity oscillations implies the RF spectrum has fewer features. To quantify the shape of the RF spectrum, we calculate the spectral flatness from the ratio between the geometric mean and the arithmetic mean of the power spectrum (see Supplementary Material). Figure~\ref{fig5}(h) shows the flatness of RF spectra integrated spatially and averaged over 5 cavities for each $g$. Its value increases by a factor of 2 in the near-planar cavities, confirming the suppression of temporal intensity oscillations of the laser emission.

\subsection{Spatio-temporal correlations}

Lastly, we compare the spatio-temporal correlations of the laser emission intensity for different cavity shapes. Figures~\ref{fig5}(i-k) show the spatio-temporal intensity correlation functions for group I, averaged over different time windows. 
 
The peak at the origin $\Delta x = \Delta t = 0$ is related to short-range correlations. Its full-width at half-maximum gives the correlation length, which reflects the average size of filaments. In the planar cavity [Fig.~\ref{fig5}(i)], the correlation length is about 6 $\mu$m, and it is halved in near-planar cavities [Figs.~\ref{fig5}(j) and (k)]. Thus the filaments become narrower when the end facets are curved. The correlation time (temporal width of the correlation function) barely changes with the cavity geometry, as it is mostly determined by the response time of the gain material (see Supplementary Material).

As the filaments move around in space and time, they induce nonlocal spatio-temporal correlations in the emission intensity~\cite{fischer1996complex,hess1996spatio,kim2021massively}. To quantify the long-range correlations, we average the magnitude of spatio-temporal correlations beyond the peak at the origin (see Supplementary Material). Figure~\ref{fig5}(l) shows the long-range spatio-temporal correlations are significantly reduced for $g$ = 0.88 and 0.74 compared to $g$ = 1. The suppression of non-local correlations indicates that the filaments are overall weaker in the near-planar cavity lasers, thus their spatio-temporal dynamics is more stable than that of the planar-cavity laser.

\section{Discussion and conclusion}

We demonstrate that the broad-area semiconductor laser characteristics can be dramatically changed by a small variation of the cavity shape. This is because the Fabry-Perot cavity with planar mirrors is located at a bifurcation point between stable and unstable ray dynamics. We curve the mirrors slightly and form a near-planar cavity with concave mirrors. As a result, the high-order transverse modes are well confined in the cavity, leading to a vast increase in the number of transverse lasing modes. The spatial coherence of laser emission is greatly reduced, which suppresses the speckle noise. Although the output beam has increased lateral divergence, its angular width is below 40$^{\circ}$. Since the lateral divergence is comparable to the vertical divergence of an edge-emitting laser, the nearly circular beam can be easily collected with standard optics. Therefore, such a laser may be used as an illumination source for full-field speckle-free imaging. The advantage of our laser compared to e.g. an incandescent lamp which also produces no speckle noise is that the lamp emits into far too many spatial modes and thus has a low power per mode, whereas our laser emits into fewer modes and thus features a higher power per mode and better directionality. The greatly improved brightness facilitates high-speed imaging through absorbing or scattering media and real-time monitoring of moving objects or transient processes. Of course, the decrease of spatial coherence will increase the focal spot size and reduce the intensity for optical pumping, material processing, and other applications. However, in these applications, not only the brightness, but also the beam profile matters, e.g., the material processing usually requires a flat-top beam, which cannot be created by tight focusing of a spatially coherent beam. On the other hand, a laser with reduced spatial coherence may directly output a flat-top beam \cite{cao2019complex}.  

Curving the end facets also leads to a drastic modification of the spatio-temporal dynamics of broad-area semiconductor lasers. With many high-order transverse modes lasing, the characteristic length scales of intensity variations in the transverse direction are greatly reduced. Consequently, the self-focusing instability induced by spatial hole burning that leads to filamentation is prevented, and the spatio-temporal instability is mitigated. For a quantitative analysis of the lasing dynamics, we develop a method to separate the intensity fluctuations caused by different processes --- filaments, mode beating, and detection noise. They have distinct spatio-temporal correlation scales, enabling us to separate filamentation from other processes. Compared to the planar cavity laser, the amplitude of spatio-temporal intensity fluctuations in the near-planar cavity is reduced by half. The RF spectrum (up to 10 GHz) becomes flattened as the temporal pulsation of emission intensity is weakened. Lastly, the reduction of filamentation in the near-planar cavity lasers decreases the long-range spatio-temporal correlations of intensity fluctuations. The stabilized laser output with negligible long-range spatio-temporal correlation will be useful for parallel random number generation~\cite{kim2021massively}.

To conclude, our method efficiently controls the nonlinear lasing dynamics by tailoring the resonator shape in the vicinity of a bifurcation point. The dramatic change in the spatial structures of cavity modes strongly affects their nonlinear interactions with the gain material. Our method is simple, robust, and works for a wide range of pump currents. It may be applied to high-power fiber and solid-state lasers, as well as other nonlinear dynamical systems. It can also be employed to control the time-reversed lasing and coherent perfect absorption~\cite{chong2010coherent,ghobadi2018strong,liu2018hybrid,tang2020plasmonic}.


\section*{Acknowledgments}
We acknowledge the computational resources provided by the Yale High Performance Computing Cluster (Yale HPC). H.C. acknowledges support by the Office of Naval Research under the Grant No. N00014-221-1-2026. S.B. acknowledges support for the Chair in Photonics from Minist\`ere de l'Enseignement Sup\'erieur, de la Recherche et de l'Innovation, R\'egion Grand-Est, D\'epartement Moselle, European Regional Development Fund (ERDF), Metz M\'etropole, GDI Simulation, CentraleSup\'elec, and Fondation CentraleSup\'elec. Q. J. Wang acknowledges support from Singapore A*STAR funding A18A7b0058, and Singapore National Research Foundation funding NRF-CRP19-2017-01 and NRF-CRP23-2019-0007. O.H. acknowledges support by the Science Foundation Ireland (SFI) via Grant No. 18/RP/6236.

\textit{Competing interests.}
The authors declare no competing financial interests.


\renewcommand{\theequation}{S\arabic{equation}}
\renewcommand{\thefigure}{S\arabic{figure}}
\renewcommand{\thetable}{S\arabic{table}}
\setcounter{figure}{0}
\setcounter{equation}{0}
\setcounter{section}{0}

\renewcommand{\thesubsection}{\arabic{subsection}} 
\renewcommand{\thesubsubsection}{\Alph{subsubsection}} 

\section*{Supplementary Material}

\section{Materials and methods}

\subsection{Numerical modeling}

The passive modes of planar and near-planar cavities are calculated with the eigenfrequency analysis module of COMSOL Multiphysics. The simulated cavities have a longitudinal length $L = $ 20 $\mu$m, and a transverse width $W = L/\sqrt{2} = 14.1$ $\mu$m. We calculate the optical modes with transverse electric (TE) polarization, which has preferential gain in the GaAs/AlGaAs quantum wells and thus dominates the lasing emission. The optical modes are the solutions of the scalar Helmholtz equation
\begin{equation} 
[\nabla^2 + k^2 n^2(x,z)] \psi(x,z) = 0 
\end{equation}
where $k$ is the wavevector and $\psi$ is the out-of-plane component $H_y$ of the magnetic field. To impose outgoing wave boundary conditions, we set a perfectly matched layer surrounding the entire simulation area, with a distance larger than 2.5 $\mu$m from the cavity boundary. For a planar cavity ($g = 1$), in order to simulate the finite width $W$ of the pumped region, perfectly matched layers are placed at the transverse boundaries ($x$ = $\pm W/2$).

\subsection{Device fabrication}

The laser cavities, of both near-planar ($g$ = 0.74 and 0.88) and planar geometries ($g$ = 1), are fabricated by etching a GaAs/AlGaAs quantum well wafer. All tested cavities have a longitudinal length $L = $ 400 $\mu$m and a lateral width $W = L/\sqrt{2} = $282 $\mu$m, but the radius of curvature $R$ of the two facets varies slightly to have different $g$. We use a commercial diode laser wafer (Q-Photonics QEWLD-808). The gain medium is a 12 nm-thick GaAs/AlGaAs quantum well. It is embedded at the middle of 0.4 $\mu$m-thick Al$_{0.37}$Ga$_{0.63}$As guiding layer, which is between p-doped and n-doped Al$_{0.55}$Ga$_{0.45}$As cladding layers with a thickness of 1.5 $\mu$m each.

The devices are fabricated with the following process. First, the bottom metal contact, made of Ni/Ge/Au/Ni/Au layers with thicknesses of 5/20/100/20/160 nm, is deposited and thermally annealed at $385\,^{\circ}\mathrm{C}$ for $30$~s. Then a mask layer, SiO$_2$ with a thickness of $300$~nm, is deposited on the front side of the wafer. Photolithography is used to define the geometry of cavities. The cavity shapes are transferred to the SiO$_2$ mask by reactive ion etching (RIE) with a mixture of CF$_{4}$ (30 sccm) and CHF$_{3}$ (30 sccm). After removing the photoresist, inductively coupled plasma (ICP) dry etching on the SiO$_2$ mask is conducted to create the cavities. Here a plasma mixture of Ar (5 sccm), Cl$_2$ (4 sccm), and BCl$_3$ (4.5 sccm) is used. The etch depth is 3.5 $\mu$m, etching through the entire guiding layer and partially into the bottom cladding layer of the wafer. The SiO$_2$ mask is removed by RIE afterwards.

After the cavity is formed, a metal contact is deposited on top of the cavity. The shapes of the top metal contacts are defined by negative photolithography. It is followed by deposition of metal contacts, made of Ti/Au layers with thicknesses $20/200$~nm, respectively. The geometry of the top contact is designed to match the spatial profile of highest-order transverse modes, in order to maximize the current injection to the lasing modes. The top contact boundaries are 6 $\mu$m away from the cavity sidewalls, in order to avoid the top contacts hanging down and blocking the emission from facets. Lastly, lift-off is performed. The sample is cleaned by O$_2$ plasma to finalize the fabrication.

\begin{figure*}[t]
\centering
\includegraphics[width=0.95\linewidth]{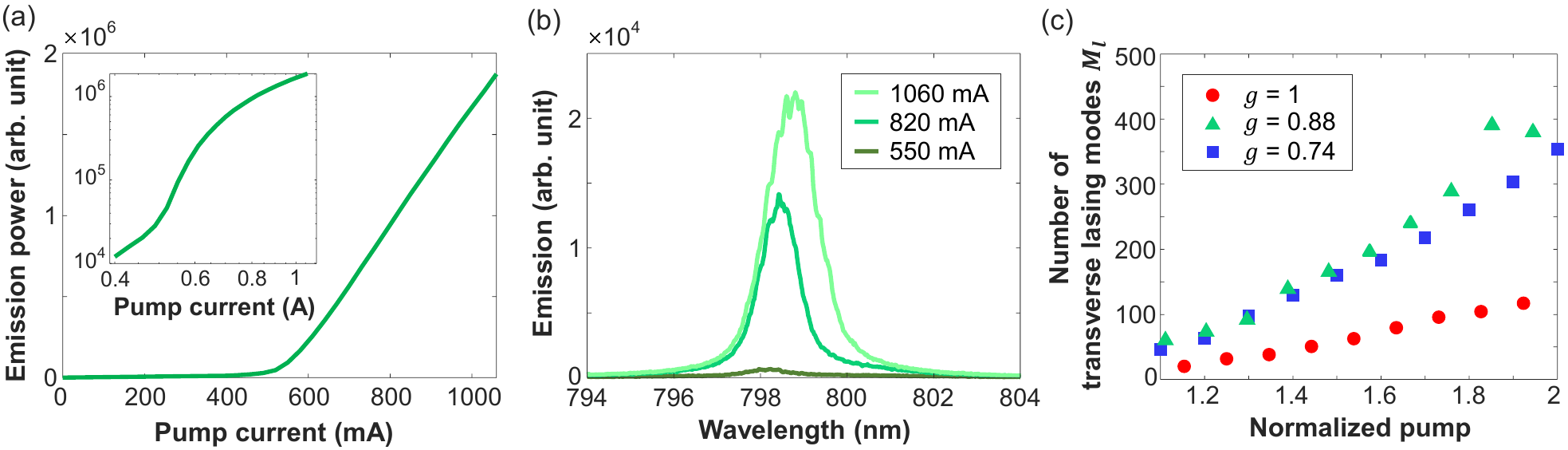}
\caption{Lasing characteristics. (a) The LI curve of a near-planar ($g = 0.88$) cavity laser shows a clear threshold at the pump current of 500 mA. Inset: the LI curve in logarithmic scale featuring the S-shape that is typical for lasers. (b) Optical spectra of lasing emission of the cavity in (a) at pump currents just above, 1.5 times, and 2 times the lasing threshold. Increasing the injection current above the threshold broadens the lasing spectrum. (c) The number of transverse lasing modes $M_l$ in the planar ($g = 1$) and near-planar ($g = 0.74, 0.88$) cavities increases with the pump level. For each cavity, the pump current is normalized by the lasing threshold.}
\label{figS1}
\end{figure*}

\subsection{Laser testing}

The device is mounted on a copper plate, and a tungsten needle (Quater Research H-20242) is placed on the top metal contact for current injection. A diode driver (DEI Scientific, PCX-7401) provides the electric current to the device. In order to minimize the device heating, we inject 2 $\mu$s long pulses at a repetition rate lower than 1 Hz. The emission is collected by a 20$\times$ microscope objective lens (NA = 0.4), and then coupled into a multimode optical fiber (diameter 600 $\mu$m, NA = 0.48). At the other end of the optical fiber, the emission is coupled into an imaging monochromator (Acton SP300i) with an intensified CCD camera (Andor iStar DH312T-18U-73) for measuring the emission spectrum.

An exemplary LI curve of a laser diode with $g = 0.88$ is shown in Fig.~\ref{figS1}(a). The typical lasing threshold current is about 500 mA. The threshold current densities, averaged for multiple cavities of $g$ = 0.74, 0.88, and 1, are $(0.52\pm0.02)~\mathrm{kA/cm^2}$, $(0.52\pm0.03)~\mathrm{kA/cm^2}$, and $(0.49\pm0.03)~\mathrm{kA/cm^2}$, respectively. Therefore, no significant difference in the lasing threshold current density is found for different cavity shapes. The optical spectrum of the lasing emission is centered at 799 nm [Fig.~\ref{figS1}(b)]. The spectral width is about 1 nm, with individual lasing peaks so densely packed that they cannot be resolved by our spectrometer.

\subsection{Spatial coherence measurement}

We measure the number of transverse lasing modes using the speckle pattern generated by our devices. The number of transverse lasing modes $M_l$ can be estimated by $1/C^2$, where the speckle contrast $C$ is the standard deviation of intensity $\sigma_I$ normalized by the mean $\langle I \rangle$ ~\cite{goodman2007speckle,redding2015low,kim2019electrically}. The speckle pattern is created by a line diffuser (RPC Photonics, EDL-20) that randomly scatters the emission only in transverse direction. In our setup, the far-field emission from the laser illuminates the line diffuser. Then we record the scattered intensity pattern in the far-field of the diffuser using a CCD camera (Allied Vision, Mako G-125B). For each laser, we repeatedly measure the speckle patterns generated at 5 different positions of the line diffuser with distinct realizations of disorder, and averaged the speckle contrasts.

Fig.~\ref{figS1}(c) shows the measured number of transverse lasing modes $M_l$ as we gradually increase the injection current up to two times of the lasing threshold. For all cavity shapes, $M_l$ increases with the pump strength, as more high-order transverse modes manage to lase. The near-planar cavities ($g = $0.74, 0.88) have larger $M_l$ than the planar cavity ($g = 1$) over the entire range of pump current measured.

\subsection{Far-field characterization}

To measure the far-field emission profile, we place a CCD camera (Allied Vision, Mako G-125B) at a distance of 0.1 m from the cavity. This distance is on the order of the Fraunhofer distance $\sim W^2/\lambda = $ 0.1 m, thus the emission is recorded in the far-field of the laser. No optics is placed between the cavity and the camera. We shift the CCD camera on a rail laterally while rotating the camera to face the laser. At every position, the lasing emission produced by a single current pulse is recorded. As the separation distance $R$ between the laser and the camera depends on the lateral position of the camera, the measured intensity is rescaled by $1/R^2$. The recorded images are combined and vertically integrated to obtain the far-field intensity distribution in the horizontal direction.

Since the measured far-field emission intensity fluctuates with the far-field angle, we smooth out the intensity profile by a moving average over $5^{\circ}$, and then calculate the full-width at half-maximum (FWHM), as shown in Fig.~3 of the main text. For comparison, we characterize the divergence angle by the second-moment width, so-called D4$\sigma$ width, defined by 4 times the standard deviation of the intensity distribution. For cavities of $g$ = 1, 0.88, and 0.74, the D4$\sigma$ full divergence angle of the far-field patterns are 37$^{\circ}$, 56$^{\circ}$, and 69$^{\circ}$, respectively.

\subsection{Near-field mapping}

To measure the near-field profile, the laser emission on the cavity facet is imaged by a 20$\times$ microscope objective (NA = 0.4) and a plano-convex lens (focal length 150 mm) onto a CCD camera (Allied Vision, Mako G-125B). To record the time-resolved near-field pattern, the emission collected with the same optical configuration is imaged onto the entrance slit of the streak camera (Hamamatsu C5680). The fast sweep unit (M5676) records time traces of emission intensities at different spatial locations.

Finally we comment on the beam parameter product, the product of the beam waist size and the far field angular spread of partially coherent light. While the low spatial coherence corresponds to a large beam parameter product, a large beam parameter product does not necessarily mean low spatial coherence. For example, a spatially coherent beam passing through an optical diffuser will spread in the far-field, making the beam parameter product large, even though the fields at all spatial locations remain coherent according to the mutual coherence function. Therefore, the beam parameter product is larger or equal to a minimal value given by the degree of spatial coherence. That is why we use the speckle intensity contrast, instead of the beam parameter product, to characterize the degree of spatial coherence of our laser emission.

\begin{figure*}[t]
\centering
\includegraphics[width=0.7\linewidth]{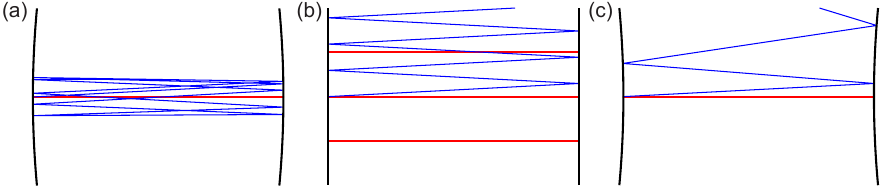}
\caption{Stability of ray dynamics in (a) a stable cavity ($g = 0.74$), (b) a Fabry-Perot cavity ($g = 1$), and (c) an unstable cavity ($g = 1.26$). The axial orbit is indicated in red, and a trajectory launched with an angular deviation of $3^\circ$ is indicated in blue. The cavities have an aspect ratio of $W/L = 1/\sqrt{2}$, and the mirrors of the cavities in (a) and (c) have a radius of curvature or $R/L \simeq 3.85$.}
\label{figS2}
\end{figure*}

\section{Ray dynamics}

As mentioned in the main text, one interesting aspect of the ray-wave correspondence in optical resonators is the influence of bifurcation points, that is, geometries at which new periodic orbits are created or change their stability when the cavity shape is varied.  Bifurcations of classical orbits have a significant effect on the properties of wave systems and require special treatment in semiclassical theories \cite{Primack1994, Schomerus1997}. For example, they lead to increased fluctuations of the density of states in microresonators \cite{Sieber1997a, Pascal2017}. Here we investigate broad-area semiconductor lasers as function of the resonator geometry in the vicinity of a bifurcation. 

We concentrate on the well-known case of a Fabry-Perot cavity with two planar facets, which is situated at the bifurcation between stable cavities with concave mirrors and unstable cavities with convex mirrors. In a stable cavity, rays in the vicinity of the axial orbit are trapped forever (assuming perfectly reflecting mirrors), while in an unstable cavity these rays will escape quickly (see Fig.~S2). Even though the passive modes change continuously with the cavity geometry, a small perturbation of the Fabry-Perot cavity can thus cause a large change in the laser properties, which is useful for sensing applications~\cite{humar2015cellular,riou2017marginally}.

\subsection{Stability}

Figure~\ref{figS2} shows examples of ray trajectories in a stable ($-1< g < 1$), a Fabry-Perot ($g = 1$) and an unstable cavity $g > 1$. It should be noted that the terms stable and unstable actually refer to the axial periodic orbit (shown in red) and how it reacts to perturbations of its initial conditions, not to the cavity itself. In general, an optical cavity can exhibit several periodic orbits with different stability \cite{Pascal2017, kim2019electrically}, hence it is misleading to call a cavity itself stable or unstable. In the following we restrict ourselves to the axial orbit which is the only relevant one here. 

A perturbation of the initial conditions can be a small change of the initial position and/or propagation direction of a trajectory. If a periodic orbit is stable, a slightly perturbed trajectory will stay close to it forever. This behavior is shown in Fig.~\ref{figS2}(a) for a cavity with $g = 0.74$: the blue trajectory, which is launched with a direction deviating by $3^\circ$ (but at the same position) always stays close to the axial orbit as it travels back and forth in the cavity. The same behavior is obtained when slightly changing the initial position. An important consequence of the stability of the axial orbit is that trajectories in its vicinity do not leave the cavity laterally and can hence support higher-order transverse modes with high $Q$ factors.

If a periodic orbit is unstable, a slightly perturbed trajectory will travel away from it at an exponential rate. This case is exemplified in Fig.~\ref{figS2}(c) for a cavity with $g = 1.26$: the blue trajectory propagates away from the axial orbit very rapidly without returning. Since the axial orbit is unstable, trajectories in its vicinity leave the cavity very quickly in the lateral direction, which greatly reduces the $Q$ factors of higher-order transverse modes. 

At the bifurcation point where the stability of the axial orbit changes from stable to unstable is the Fabry-Perot cavity ($g =1$), for which the axial orbit is marginally stable. For marginally stable orbits, different perturbations of the initial conditions yield different results. When the initial direction of the axial orbit is changed, the perturbed trajectory propagates away linearly in time and leaves the cavity after a few round trips as shown by the blue trajectory in Fig.~\ref{figS2}(b). This reduces the lifetime of higher-order transverse modes in the Fabry-Perot cavity. In contrast, perturbations of the initial position (but not direction) yield again periodic orbits as exemplified by the three red orbits in Fig.~\ref{figS2}(b). So the axial orbit is part of a family of periodic orbits, consisting of all trajectories perpendicular to the planar end mirrors. In contrast, the axial orbit is called isolated when it is stable or unstable since it is the only periodic orbit in its vicinity in these cases. 

Since the Fabry-Perot cavity is at a bifurcation point, small changes of the cavity geometry will strongly impact the behavior of trajectories around the axial orbit and lead to significant changes in lasing behavior because the stability of the axial orbit is related to the $Q$ factors of higher-order transverse modes.

\subsection{Ray trajectories and transverse modes}

\begin{figure*}[t]
\centering
\includegraphics[width=0.8\linewidth]{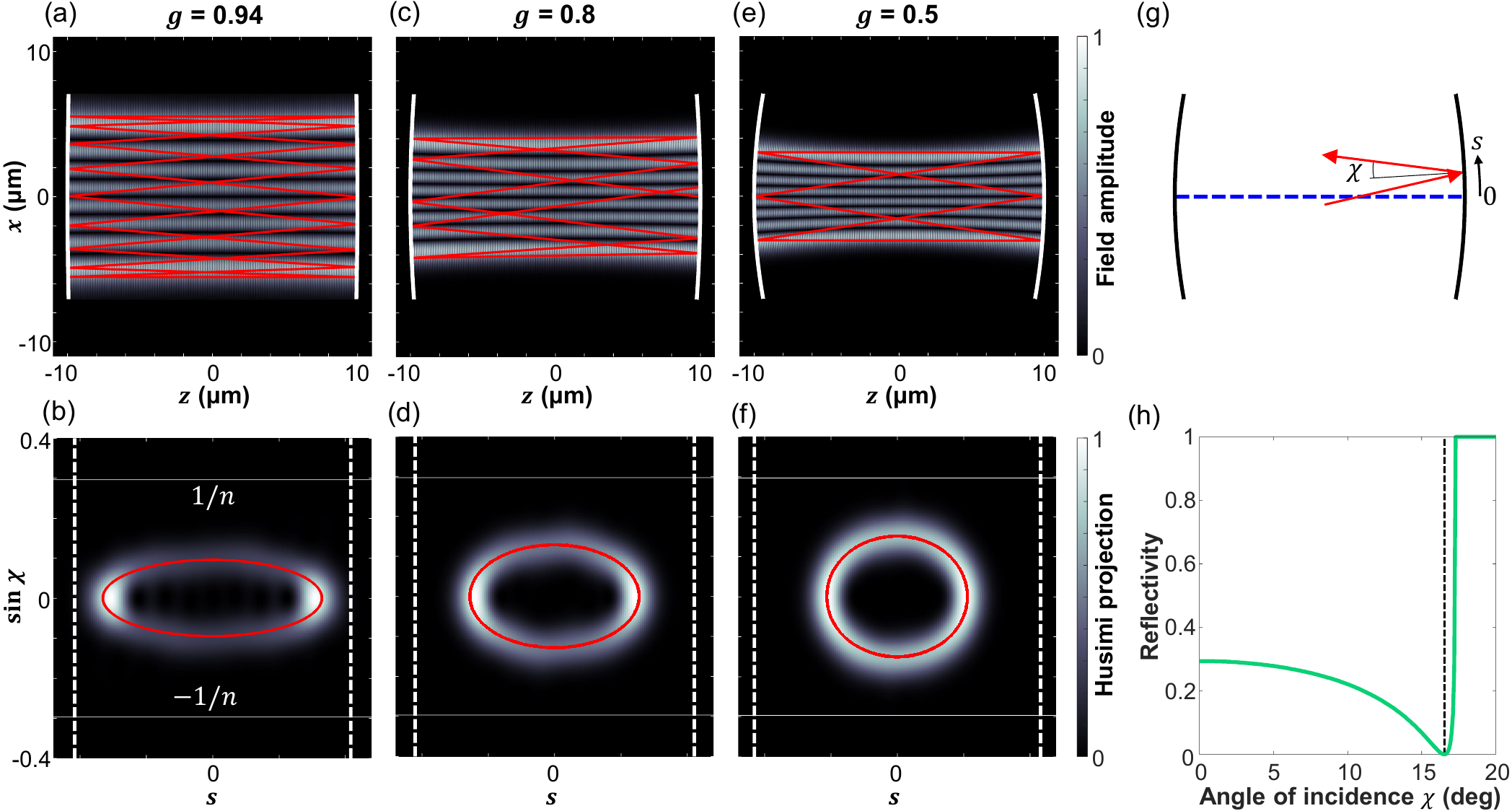}
\caption{Comparison of transverse modes of order $m = 7$ in $L = 20~\mu$m long cavities with (a, b) $g = 0.94$, (c, d) $g = 0.8$, and (e, f) $g = 0.5$. (a, c, e) Field intensity profiles in real space with (a) $\lambda = 798.4$~nm and $Q = 420.8$, (c) $\lambda = 799.8$~nm and $Q = 410.8$, and (e) $\lambda = 800.0$~nm and $Q = 399.8$. Exemplary parts of the corresponding trajectories are superimposed as red lines. (b, d, f) Husimi distributions of the modes shown in panels (a, c, e), respectively. The corresponding trajectories in Birkhoff coordinates are superimposed as red points. The vertical white dashed lines indicate the endpoints of the right mirror and the horizontal gray lines indicate the critical angle. (g) Definition of the Birkhoff coordinates: the position $s$ on the right mirror and the angle of incidence $\chi$. The blue dashed line indicates the optical axis ($s = 0$). (h) Reflectivity for p-polarized light at a semiconductor-to-air interface with refractive index contrast $3.37$. The vertical dashed lines marks the Brewster angle at $16.5^\circ$.}
\label{figS3}
\end{figure*}

In the following we compare the same transverse mode of order $m = 7$ in cavities with three different stability parameters, in order to evaluate the effect of mirror curvature on the transverse mode profile and the $Q$ factor. The real-space intensity distributions for $g = 0.94, 0.80$ and $0.50$ are shown in Figs.~\ref{figS3}(a, c, e), respectively. With decreasing $g$ (increasing mirror curvature), the transverse mode profile becomes narrower, indicating modes with increasingly high order can be confined in the cavities. However, the $Q$ factor decreases from $Q = 420.8$ for $g = 0.94$ to $Q = 399.8$ for $g = 0.5$ (cf.\ Table~\ref{tabModesRays}). The dependency of the mode width and the $Q$ factor on $g$ can be explained by analyzing the corresponding ray trajectories.

A useful tool for analyzing the ray-wave correspondence in optical microcavities are the so-called Husimi functions \cite{Husimi1940, Hentschel2003}, which can be considered as a phase-space representation of a mode. They are obtained by calculating the overlap of the wave function with a wave packet of minimal uncertainty with specific position and momentum. Here we use the Husimi function for dielectric resonators introduced in Ref.~\cite{Hentschel2003}. The Poincar\'{e} surface of section is given in Birkhoff coordinates $(s, \sin \chi)$ as shown in Fig.~\ref{figS3}(g): $s$ denotes the location of incidence of a ray on the right mirror (normalized by the total circumference of the resonator, $s_\mathrm{t}$), where $s = 0$ is at the center of the mirror, and $\chi$ is the angle of incidence with respect to the surface normal. 

The Husimi functions of the three modes considered here are shown in Figs.~\ref{figS3}(b, d, f). They have a roughly elliptic structure, where the semi-axes in $s$ direction decrease with increasing $g$ since the modes become narrower. The semiaxes in $\sin\chi$ direction increase, which means that the modes contain wave components with higher angles of incidence. 
Table~\ref{tabModesRays} gives the angles $\chi_\mathrm{max}$ at which the maxima in the $s = 0$ section of the Husimi distributions are found. The increase of $\chi_\mathrm{max}$ with decreasing $g$ can explain the decline of the $Q$ factors since the reflectivity at the semiconductor-air interface decreases towards the Brewster angle for p-polarization as shown in Fig.~\ref{figS3}(h). 

\begin{table}[b]
\begin{center}
\begin{tabular}{c|ccc|cc}
\hline \hline
$g$ & $\lambda$ (nm) & $Q$ & $\chi_\mathrm{max}$ (deg) & $\tau (L/c)$ & $Q_\mathrm{ray}$ \\
\hline
$0.94$ & $798.4$ & $420.8$ & $5.5$ & $2.68$ & $421.2$ \\
$0.80$ & $799.8$ & $410.8$ & $7.4$ & $2.62$ & $411.2$ \\
$0.50$ & $800.0$ & $399.8$ & $8.7$ & $2.56$ & $402.2$ \\
\hline \hline
\end{tabular}
\caption{Properties of the passive cavity modes and corresponding ray trajectories shown in Fig.~\ref{figS3} for different stability parameters $g$. $\lambda$ and $Q$ are the resonance wavelength and the quality factor, respectively, and $\chi_\mathrm{max}$ is the maximal angle of incidence in the Husimi function. $\tau$ is the fitted lifetime of the trajectory and $Q_\mathrm{ray}$ the corresponding quality factor for a $L = 20~\mu$m long cavity at $\lambda = 800$~nm.}
\label{tabModesRays}
\end{center}
\end{table}

\begin{figure*}[ht]
\centering
\includegraphics[width=0.7\linewidth]{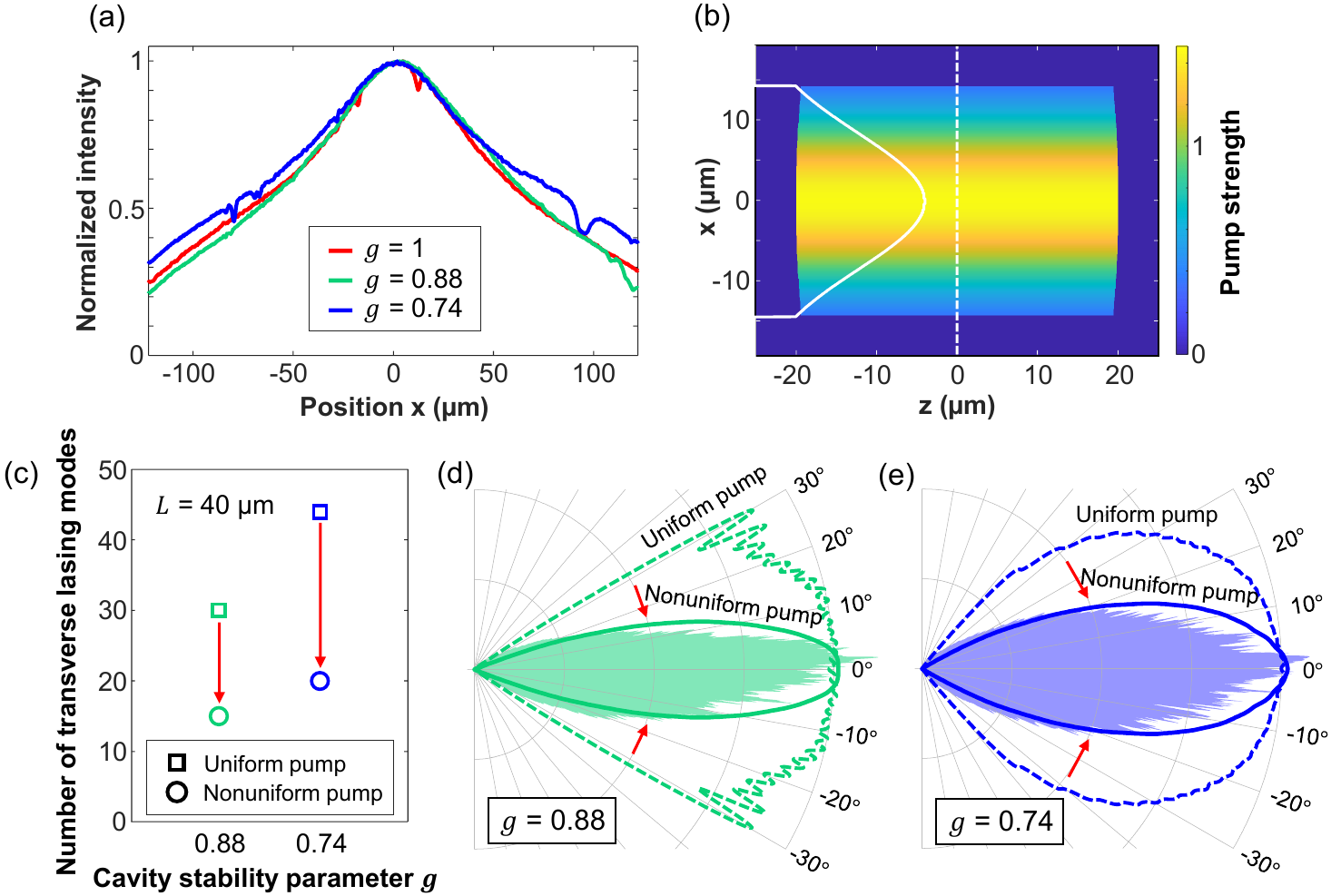}
\caption{Spatially nonuniform pumping. (a) At the pumping level of 0.4 times of the lasing threshold, the spontaneous emission dominates. Its nonuniform intensity distribution at the cavity facet reflects the nonuniform current density profile. Each curve is averaged over 5 cavities of the same geometry $g$. The field of view covers 90\% of the entire facet. (b) Simulated spatial profile of pump strength, which varies only in the transverse direction ($x$). The cross-section of the pump is shown as white line. The cavity length $L$ is 40 $\mu$m and the width $W$ is $L/\sqrt{2}$ = 28.3 $\mu$m. (c) The number of transverse lasing modes, calculated by SPA-SALT, for spatially uniform pumping (square) and nonuniform pumping (circle). The red arrows mark the drop in the number of transverse lasing modes due to spatially nonuniform pumping. (d, e) Narrowing of the far-field intensity profiles caused by spatially nonuniform pumping in near-planar cavities of (d) $g = 0.88$ and (e) $g = 0.74$. The shaded area indicates the experimentally measured far-field pattern in Fig.~3 of the main text, and the solid line represents the smoothed profile.
}
\label{figS4}
\end{figure*}

We calculate the corresponding trajectories by launching rays at the center of the mirror ($s = 0$) with the angles $\chi_\mathrm{max}$ given in Table~\ref{tabModesRays}. Their time evolution for $2000$ reflections is computed. The real space and phase space representations of these trajectories are superimposed in red in Figs.~\ref{figS3}(a-f). Each reflection at the right mirror is indicated as a point in phase space [Figs.~\ref{figS3}(b, d, f)], though the points are so dense that they appear as a continuous line. The agreement with the Husimi functions is excellent, demonstrating that these are indeed the trajectories on which the modes are based. In real space [Figs.~\ref{figS3}(a, c, e)], only the first few reflections are shown for better visibility.

The trajectories cover a finite transverse region of the cavity, and the outmost segments of the trajectories coincide with the most intense regions of the wave functions. Furthermore, the maximal angles of incidence $\chi$ of the trajectories (at the center of the mirror) increase with decreasing $g$. The intensity of a trajectory decreases at each reflection according to the Fresnel reflection coefficients for $p$ polarization [cf.\ Fig.~\ref{figS3}(h)], leading to an exponential decay in time \cite{Bittner2020}. The fitted lifetimes $\tau$ as well as the corresponding quality factors $Q_\mathrm{ray}$ for a $L = 20~\mu$m long cavity are given in Table~\ref{tabModesRays} and show excellent agreement with the $Q$ factors of the modes. This demonstrates that refraction at the semiconductor-air interfaces is the dominant loss mechanism responsible for the decrease of the $Q$ factors as the mirror curvature increases. The slightly smaller $Q$ factors of the modes can be attributed to diffraction losses not considered in the ray tracing simulations.

Therefore, maximizing the number of transverse lasing modes requires balancing two effects. On one hand, increasing the mirror curvature reduces the width of the modes so transverse modes of higher order can fit into the cavity laterally. This leads to the strong increase of the number of lasing modes when transitioning from a Fabry-Perot cavity with $g = 1$ to a stable cavity with $g < 1$. On the other hand, the refractive losses for high-order modes of p-polarization grow with the mirror curvature, leading to an increase of the lasing threshold. Both effects are intuitively explained by the corresponding ray trajectories as shown above. It should furthermore be noted that the optimal stability parameter $g$ depends both on the aspect ratio $W/L$ of the cavity and the refractive index.

\section{Spatially nonuniform pumping}

Figure~2(b) of the main text shows a similar number of transverse lasing modes $M_l$ for $g = 0.88$ and $g = 0.74$. This experimental result deviates from the simulation which predicts a larger $M_l$ for $g = 0.74$ in Fig.~1(e) of the main text. The numerical simulation assumes spatially uniform pump. Experimentally, the current injection through the top contact may be nonuniform, which will modify $M_l$. To resolve the discrepancy of $M_l$, we investigate spatially nonuniform pumping in this section. 

\subsection{Spatial mapping of current injection}

To map the spatial profile of current injection, we measure the emission intensity distribution on the cavity facet at a pumping level well below the lasing threshold. The spontaneous emission dominates over the stimulated emission, and its spatial profile directly reflects the current distribution. Figure~\ref{figS4}(a) shows the near-field emission patterns for $g$ = 1, 0.88 and 0.74. The emission spectra do not change with the pump current, confirming that stimulated emission is negligible. In all three cavities, the spontaneous emission is stronger in the center, reflecting larger current density there than close to the boundaries.

\subsection{Number of transverse lasing modes}

To analyze how the spatially nonuniform pumping affects lasing, we use the SPA-SALT (single-pole approximation steady-state \textit{ab-initio} theory)~\cite{ge2010steady,liew2015pump,cerjan2016controlling} to calculate the number of lasing modes and the modal intensities. Simulating the experimental cavities of length $L$ = 400 $\mu$m and width $W$ = 283 $\mu$m is computationally demanding. Therefore we reduce the cavity dimension to $L$ = 40.0 $\mu$m and $W$ = 28.3 $\mu$m, but keep the aspect ratio $L/W = \sqrt{2}$. 

With spatially uniform pumping, the number of transverse lasing modes at a pumping level of two times the lasing threshold are 44 and 30 [squares in Fig.~\ref{figS4}(c)] for $g = 0.74$ and $g = 0.88$, respectively. We note that these numbers are twice of those in Fig.~1(e) of the main text for cavities of half the size, $L$ = 20 $\mu$m. Thus, the number of transverse lasing modes increases linearly with the cavity size.

Here we apply the spatially nonuniform pumping in the transverse direction. As shown in Fig.~\ref{figS4}(a), the pump strength is maximal at the center and decays towards the lateral boundaries. The transverse pump profile in simulations [Fig.~\ref{figS4}(b)] is approximated by a Gaussian function with its full width at half maximum of 19.8 $\mu$m, which is 0.7 times the cavity width $W$. Such a profile promotes lasing in the lower order transverse modes which are concentrated in the center and have larger overlap with the pump than the higher order transverse modes. As the lower order modes lase efficiently and saturate the optical gain, it is harder for higher-order modes to lase. Consequently, the number of transverse lasing modes $m_l$ decreases [circles in Fig.~\ref{figS4}(c)].

The reduction in the number of transverse lasing modes is larger for $g = 0.74$ than $g = 0.88$. This is attributed to the different strength of lateral confinement of lasing modes, which affects their competition for gain. The $g = 0.74$ cavity has stronger lateral confinement and supports a larger number of transverse modes than $g = 0.88$. The transverse nonuniform pumping further enhances the mode competition, as the low-order transverse modes become dominant and more effectively prevent the high-order transverse modes from lasing. This effect is more severe in $g = 0.74$, which has a larger number of high-order transverse modes and tighter lateral confinement than $g = 0.88$. It leads to a larger decrease of the number of transverse lasing modes for $g = 0.74$. As a result, the number of transverse lasing modes of $g = 0.74$ becomes more similar to that of $g = 0.88$. 

\subsection{Far-field divergence}

The far-field emission patterns are also affected by the spatially nonuniform pumping. As low-order transverse modes start to dominate lasing, the divergence of far-field emission becomes narrower. We calculate the far-field intensity patterns for near-planar cavities with $L$ = 40 $\mu$m. Figures~\ref{figS4}(d) and (e) show narrowing of the far-field emission once the pump becomes spatially nonuniform. The divergence angle (full width at half maximum) drops from 63$^{\circ}$ to 29$^{\circ}$ for $g$ = 0.88, and from 82$^{\circ}$ to 40$^{\circ}$ for $g$ = 0.74.

\section{Spatio-temporal dynamics}

\subsection{Filament-induced fluctuations}

As discussed in the main text, the measured intensity fluctuations are caused by three distinct processes: (i) filamentation and pulsation, (ii) spatio-temporal beating of lasing modes, and (iii) photo-detection noise of the streak camera. We apply singular value decomposition (SVD) on the streak camera images to separate filament-induced fluctuations from the other two. Below we elaborate on the separation procedure based on distinct spatial and temporal scales of these fluctuations.

We first acquire an ensemble of spatio-temporal intensity distributions from the same laser cavity. Within a 2-$\mu$s-long pump current pulse, we measure 161 consecutive 10 ns-long streak camera images, which constitute a 1.61 $\mu$s-long total time window. The transient regime at the beginning of the current pulse is excluded. We denote the intensity fluctuations $\delta I^{(i)}(x,t)$ with the superscript $i = 1,...,161$ to indicate a series of streak camera images acquired at 161 measurement start times $t^{(i)}$. The difference between consecutive start times $t^{(i+1)}-t^{(i)}$ is 10 ns, equal to the duration of the individual streak images. The SVD of $\delta I^{(i)}(x,t)$ yields the singular values $s^{(i)}_{\alpha}$ and the spatial and temporal singular vectors $u^{(i)}_{\alpha}(x)$ and $v^{(i)}_{\alpha}(t)$, where $\alpha$ is the SVD index.

\begin{figure*}[htbp]
\centering
\includegraphics[width=0.8\linewidth]{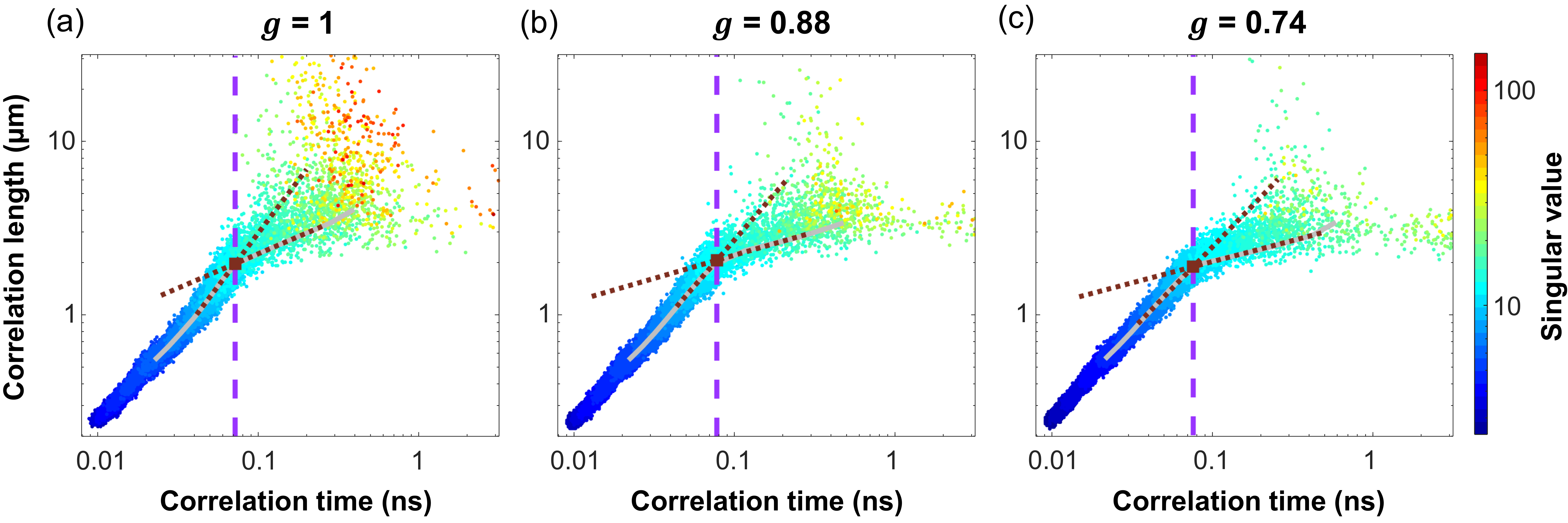}
\caption{Spatial and temporal correlation scales of singular vectors. Scatter plot of correlation length versus correlation time for all singular vectors in 161 time windows for the three cavity geometries (a) $g = 1$, (b) $g = 0.88$, and (c) $g = 0.74$. The color of each data point reflects the magnitude of the singular value. The gray solid lines represent the average of all correlation times with the same correlation length. A linear fit (brown dotted lines) defines the correlation time (vertical dashed purple line) at which the slope of the gray line changes suddenly. It separates group I, singular vectors due to filaments, from groups II and III. }
\label{figS5}
\end{figure*}

The characteristic length scales of the spatial and temporal singular vectors are given by the full-width at half-maximum (FWHM) of the correlation function given in Equation (2) of the main text. Figure~\ref{figS5} is a scatter plot of the correlation length $l^{(i)}_{\alpha}$ vs. the correlation time $\tau^{(i)}_{\alpha}$ of every singular vector of index $\alpha$ in the time window $t^{(i)}$. The gray line denotes the averaged correlation time at a fixed correlation length. For all cavity geometries, these lines exhibit a sharp change in slope. In order to define this transition, linear fitting is applied to find its exact position, which gives the critical correlation time $\tau_c$. Typical $\tau_c$ is between 70 ps and 80 ps, and there is no systematic dependence of $\tau_c$ on the cavity geometry $g$. Group I includes all singular vectors with correlation times exceeding $\tau_c$ [on the right side of the dashed purple line in Fig.~\ref{figS5}]. For the $i$-th time window, the group I of singular vectors is given by:
\begin{equation}
\mathrm{I}^{(i)} := \{ \alpha | \tau^{(i)}_{\alpha} > \tau_c \}.
\end{equation}
The spatio-temporal intensity fluctuation $\delta I_\mathrm{I}^{(i)}(x,t)$ caused by filaments is reconstructed by summing the singular vectors in group I,
\begin{equation}
\delta I_\mathrm{I}^{(i)}(x,t) = \sum_{\alpha \in \mathrm{I}^{(i)}} s^{(i)}_{\alpha} u^{(i)}_{\alpha}(x) v^{(i)}_{\alpha}(t).
\end{equation}

\subsection{Fluctuation amplitude}

Figure~\ref{figS5} shows the magnitude of singular values by color. The planar cavity $g = 1$ has a larger number of singular vectors with higher singular values. Here the singular value represents the fluctuation power carried by the corresponding spatio-temporal singular vector, as described below.

The SVD of the spatio-temporal intensity fluctuation $\delta I(x,t)$, of dimension $N_x \times N_t$, is given by [the superscript $(i)$ is dropped for simplicity],
\begin{equation}
\delta I(x,t) = \sum_{\alpha} s_{\alpha} u_{\alpha}(x) v_{\alpha}(t),
\label{eqS_sv1}
\end{equation}
where $s_{\alpha}$ is the singular value with index $\alpha$, $u_{\alpha}$ is the coresponding spatial singular vector, and $v_{\alpha}$ the temporal singular vector. The singular vectors are ortho-normal:
\begin{equation}
\langle u_{\alpha}(x) u_{\beta}(x) \rangle_x = \delta_{\alpha\beta}/N_x
\label{eqS_sv2}
\end{equation}
\begin{equation}
\langle v_{\alpha}(t) v_{\beta}(t) \rangle_t = \delta_{\alpha\beta}/N_t
\label{eqS_sv3}
\end{equation}
where $\delta_{\alpha\beta}$ is the Kronecker delta. The fluctuation power of the spatio-temporal intensity is given by the variance of the intensity fluctuation $\langle \delta I(x,t)^2 \rangle_{x,t}$, where the average fluctuation $\langle \delta I(x,t) \rangle_{x,t} =0$. Using the definition of SVD in Equation~(\ref{eqS_sv1}), the fluctuation power is
\begin{equation}
\begin{split}
\langle \delta I(x,t)^2 \rangle_{x.t} & = \langle \{ \sum_{\alpha} s_{\alpha} u_{\alpha}(x) v_{\alpha}(t) \}^2 \rangle_{x,t} \\
& = \sum_{\alpha,\beta}  \langle s_{\alpha} u_{\alpha}(x) v_{\alpha}(t) s_{\beta} u_{\beta}(x) v_{\beta}(t) \rangle_{x,t} \\
& = \sum_{\alpha,\beta}  s_{\alpha} s_{\beta} \langle u_{\alpha}(x) u_{\beta}(x) \rangle_x \langle v_{\alpha}(t) v_{\beta}(t) \rangle_t.
\end{split}
\label{eqS_sv4}
\end{equation}
Using the orthonormality of the singular vectors [Equations~(\ref{eqS_sv2}) and (\ref{eqS_sv3})],
\begin{equation}
\begin{split}
\langle \delta I(x,t)^2 \rangle_{x.t} & = \frac{1}{N_x N_t} \sum_{\alpha,\beta} s_{\alpha} s_{\beta} \delta_{\alpha\beta} \\
& = \frac{1}{N_x N_t} \sum_{\alpha} s_{\alpha}^2 \, 
\end{split}
\label{eqS_sv5}
\end{equation}
where $N_x$ and $N_t$ are the total numbers of spatial and temporal sampling points in a streak image. This result shows that the singular value squared $s_\alpha^2$ represents the contribution of the $\alpha$-th singular vector to the spatio-temporal intensity fluctuation.

The fluctuating power $S^{(i)}$, carried by the filaments $\delta I^{(i)}_\mathrm{I}(x,t)$ in a time window $i$, can be written as
\begin{equation}
S^{(i)} = \langle \{ \delta I^{(i)}_\mathrm{I}(x,t) \}^2 \rangle_{x,t} = \frac{1}{N_x N_t}\sum_{\alpha \in \mathrm{I}^{(i)}} [ s^{(i)}_{\alpha} ]^2,
\end{equation}
This relation indicates that the intensity fluctuation caused by filaments can be described by summing over the square of singular values belonging to group I. The fluctuation amplitude $p$ is hence defined as
\begin{equation}
p = \sqrt{\langle S^{(i)} \rangle_i},
\end{equation}
where $\langle \cdot \rangle_i$ is the average over different time windows.

\begin{figure*}[htbp]
\centering
\includegraphics[width=0.7\linewidth]{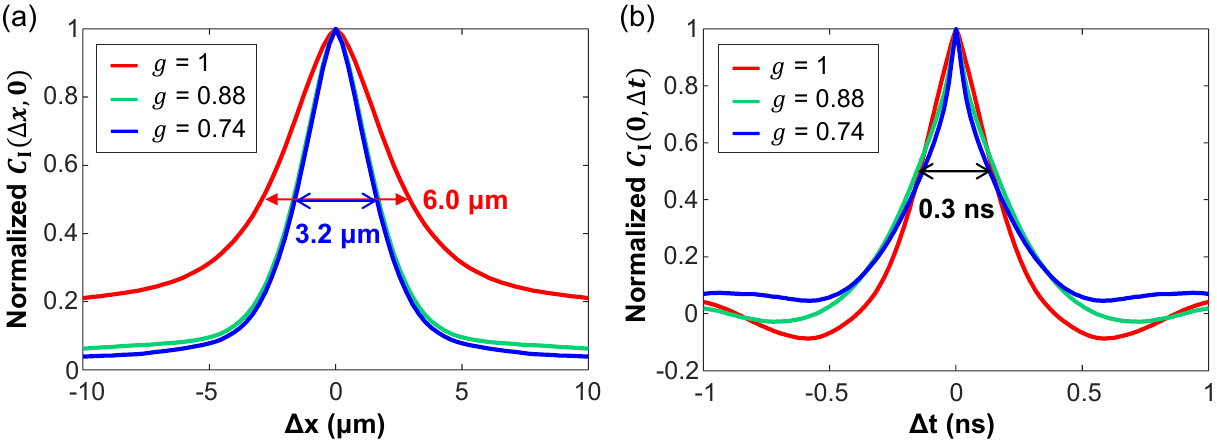}
\caption{(a) Spatial correlation function $C_{\mathrm{I}}(\Delta x,0)$ and (b) temporal correlation function $C_{\mathrm{I}}(0, \Delta t)$ of intensity fluctuations caused by filaments in different cavity geometries. The correlation functions are averaged over 5 different cavities for each value of $g$, to account for device-to-device variations. The FWHM of the correlation functions is indicated with arrows. The spatial correlation width decreases by a factor of 2 for the near-planar cavities ($g = 0.88$, 0.74), whereas the temporal correlation width remains nearly identical to that for the planar cavity ($g = 1$).}
\label{figS6}
\end{figure*}

\subsection{RF spectrum}

We compute the radio-frequency (RF) spectrum of the intensity fluctuations caused by filaments. The spatially-resolved RF spectrum, shown in Figs.~5(e)-(g) of main text, is given by,
\begin{equation}
P(x,f) = \langle |\mathcal{F}\{\delta I_{\mathrm{I}}^{(i)}(x,t)\}|^2 \rangle_{i}
\label{eqS_RF}
\end{equation}
where $\mathcal{F}$ is the Fourier transform in time. We emphasize that the averaged intensity $\langle I^{(i)}(x,t) \rangle_{x,t}$ is normalized to 1 for all cavities in order to compare the RF power. 

Before calculating the RF spectrum flatness, we spatially integrate the RF spectrum $\langle P(x,f) \rangle_x$. The RF spectrum flatness $F_s$ is defined by the geometric mean divided by the arithmetic mean in a frequency domain,
\begin{equation}
F_s = \frac{\mathrm{exp}{\big\{ \langle \mathrm{ln} [ \langle P(x,f) \rangle_x  ] \rangle_f \big\} }}{\langle P(x,f) \rangle_{x,f}}.
\end{equation}
As the filament-induced intensity fluctuations are on the order of a few GHz, the flatness is computed within the frequency range up to 10 GHz.

\subsection{Spatio-temporal correlations}

The spatio-temporal correlation of laser intensity fluctuations caused by filaments (group I) is defined by, 
\begin{equation}
\begin{split}
C_{\mathrm{I}}(\Delta x,\Delta t) & = \langle C_{\mathrm{I}}^{(i)}(\Delta x,\Delta t) \rangle_i \\
& = \langle \delta I^{(i)}_\mathrm{I}(x,t)\delta I^{(i)}_\mathrm{I}(x+ \Delta x,t+ \Delta t)\rangle_{x,t,i}.
\end{split}
\end{equation}
It is peaked at the origin $\Delta x =0$, $\Delta t = 0$. This peak represents the local correlations of intensity fluctuations, and its width gives the characteristic scale of the filaments. Beyond this peak are the nonlocal spatio-temporal correlations introduced by transverse movement and temporal pulsation of filaments~\cite{fischer1996complex}.

Figure~\ref{figS6}(a) shows the spatial correlation function $C_{\mathrm{I}}(\Delta x,0)$ for different cavity geometries. Its FWHM gives the correlation length. The planar cavity  ($g = 1$) has a correlation length of 6.0 $\mu$m, while in the near-planar cavities ($g$ = 0.88 and 0.74) the correlation length is reduced by a factor of 2 to 3.2 $\mu$m. It reflects the decrease of filament size, as more high-order transverse modes lase in the near-planar cavity.   Moreover,  $C_{\mathrm{I}}(\Delta x,0)$ for $g = 1$ exhibits a long tail at large $\Delta x$, reflecting  the long-range correlations induced by filaments. In the near-planar cavities, the long tail of $C_{\mathrm{I}}(\Delta x,0)$ is removed as a result of filament suppression.  

Figure~\ref{figS6}(b) shows the temporal correlation function $C_{\mathrm{I}}(0,\Delta t)$, averaged over 5 different cavities for each $g$. Its width gives the correlation time, which is approximately 0.3 ns, with little dependence on the cavity geometry. It is dictated by the inherent response time of carrier dynamics in the GaAs quantum well. For $g$ = 1, the negative correlation at $|\Delta t|$  = 0.5 ns is more pronounced than for $g$ = 0.74 or 0.88, reflecting stronger long-range temporal correlations in the planar cavity.

To examine the nonlocal spatio-temporal correlations, we separate the regions of short-range and long-range correlations in Figs.~5(i-k) of the main text. Their boundary is set by an ellipse whose semi-axes are the spatial and temporal FWHM of the peak at origin. Then we average the modulus of $C_{\mathrm{I}}(\Delta x,\Delta t)$ outside the ellipse but within the range of $|\Delta x| < $ 30 $\mu$m and $|\Delta t| < $  1 ns, where most correlations exist.

\end{document}